\documentclass{AM}
\usepackage{float}
\usepackage{dcolumn,graphicx,color,booktabs,microtype,afterpage}
\usepackage{amssymb}
\usepackage{amsmath}
\usepackage[charter,greekuppercase=italicized]{mathdesign}
\usepackage[charter]{mathdesign}
\usepackage{sidecap}
\usepackage[mathlines]{lineno}
\usepackage[numbers,sort&compress]{natbib}

\graphicspath{{./}{figure/}}
\renewcommand{\tablename}{Table}
\makeatletter\renewcommand{\fnum@figure}[1]{\figurename~\thefigure.~}\makeatother
\makeatletter\renewcommand{\fnum@table}[1]{\tablename~\thetable.}\makeatother

\newcount\hh \newcount\mm
\hh=\time \divide\hh by 60
\mm=\hh \multiply\mm by 60 \mm=-\mm
\advance\mm by \time
\def\now{\number\hh:\ifnum\mm<10{}0\fi\number\mm}

\usepackage[colorlinks,plainpages=false,linkcolor=blue,urlcolor=blue,citecolor=blue,pdfpagemode=UseNone,pdfstartview=FitBH]{hyperref}

\newcommand{\tcr}[1]{\textcolor{black}{#1}}

\UseRawInputEncoding

\begin{document}
	
\pagestyle{fancy}
\rhead{\includegraphics[width=2.5cm]{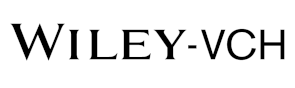}}	
%
%
\title{Unconventional superconductivity in ScIr$_2$ chiral crystal with a kagome lattice}

\maketitle

\author{Keqi Xia$\dag$}
\author{Jianzhou Zhao$\dag$}
\author{Igor Plokhikh$\dag$}
\author{Marisa Medarde}
\author{Yang Xu}
\author{Qingfeng Zhan}
\author{Dariusz Jakub Gawryluk*}
\author{Toni Shiroka*}
\author{Tian Shang*}

\dedication{\dag These authors contributed equally to this work}

%
\begin{affiliations}
	

\vspace{4mm}
K. Xia, Prof. Y. Xu, Prof. Q. Zhan, Prof. T. Shang\\
School of Physics, East China Normal University, Shanghai 200241, China\\
Email Address: tshang@phy.ecnu.edu.cn\\	

\vspace{4mm}
Prof. Y. Xu, Prof. Q. Zhan,  Prof. T. Shang\\
Key Laboratory of Polar Materials and Devices (MOE), East China Normal University, Shanghai 200241, China\\

\vspace{4mm}
Prof. J. Zhao\\
Department of Physics, School of Science, Tianjin University, Tianjin, 300354, China\\

\vspace{4mm}
Dr. I. Plokhikh, Dr. M. Medarde, Dr. D. Gawryluk, Dr. T. Shiroka\\ 
PSI Center for Neutron and Muon Sciences CNM, CH-5232 Villigen PSI, Switzerland\\
Email Address: dariusz.gawryluk@psi.ch

\vspace{4mm}
Dr. I. Plokhikh\\
Technical University Dortmund, Physics Department, Otto-Hahn-Str. 4, 44227 Dortmund, Germany\\

\vspace{4mm}
Dr. T. Shiroka\\
Laboratorium f\"ur Festk\"orperphysik, ETH Z\"urich, CH-8093 Z\"urich, Switzerland\\
Email Address: tshiroka@phys.ethz.ch
	
\end{affiliations}


\vspace{4mm}
\keywords{kagome lattice, spin-orbit coupling, unconventional superconductivity, topological chiral crystal}

\begin{abstract}
\justifying

Materials with a kagome lattice host exotic quantum phenomena driven by the interplay between band topology, spin-orbit coupling, magnetism, and electronic correlations. While magnetism of kagome materials has been widely investigated, their unconventional superconductivity (SC) 
remains largely unexplored due to the limited availability of suitable materials. Here, we report evidence of unconventional SC in the ScIr$_{2-x}$Si$_{x}$ family by combining muon-spin spectroscopy measurements with band-structure calculations. The parent ScIr$_2$ undergoes a structural phase transition from a high-$T$ cubic- to a low-$T$ rhombohedral phase, while the Ir kagome layer remains, albeit slightly  distorted.
Although the structural transition is suppressed by Si substitution, the superconducting pairing of ScIr$_{2-x}$Si$_{x}$ remains well described by a two-gap model. Since at least one of the gaps has nodes, this indicates an unconventional SC.
Its unconventional nature can be explained by the distinct flat bands occurring near the Fermi level, leading to strong electronic correlations in
the ScIr$_{2-x}$Si$_{x}$ family. Moreover, the low-$T$ phase of ScIr$_2$ exhibits an Ir chiral chain; therefore, it can be classified as a topological chiral crystal. Overall, the unusual properties of the ScIr$_{2-x}$Si$_{x}$ family make it an interesting, albeit rare, system for studying the interplay between unconventional SC, flat bands, and chirality.
\end{abstract}
\clearpage

\justifying
\section{Introduction}
Kagome lattice, a geometrically frustrated network of corner-sharing triangles, has become one of the hottest topics in condensed matter physics. Materials with a kagome lattice harbor nontrivial electronic structures generated by its unique geometry, including Dirac fermionic excitations, dispersionless flat bands, van Hove singularities (VHSs), etc. Flat bands arise from the destructive quantum interference of localized states which, together with the VHSs, imply strong electronic correlations. In kagome materials, the interplay of band topology, spin-orbit coupling (SOC), electronic correlations, and magnetism leads to rich and exotic quantum phenomena~\cite{Yin2022,Neupert2022,Wang2023,Guguchia2023,Jiang2023,Checkelsky2024,Wilson2024,He2025,Wang2025,Negi2025}, such as, superconductivity (SC), charge-density waves (CDW), spin-density waves (SDW), topological spin textures, quantum spin liquid (QSL), etc. 

Magnetic kagome materials that exhibit magnetic frustration, QSL states~\cite{Zhou2017,Savary2017,Broholm2020,Balents2010}, or anomalous Hall effect~\cite{Yin2022,Wang2023,Negi2025,Nakatsuji2015,Liu2018} have been widely studied. However, their superconducting counterparts, in particular unconventional, remain largely unexplored due to the limited availability of suitable kagome materials. Although different families of superconducting kagome materials have been experimentally discovered or theoretically predicted, most of them either behave as conventional superconductors or have not yet been synthesized.
One of the exceptions is the $A$V$_3$Sb$_5$ ($A$ =  K, Rb, Cs) family~\cite{Wilson2024,Ortiz2019,Ortiz2020}, which becomes superconducting below $T_c$ $\approx$ 0.9-2.5\,K and exhibits a variety of exotic properties, including unconventional SC~\cite{Mielke2022b,Guguchia2023b}, chiral CDW order~\cite{Jiang2021}, pair-density wave~\cite{Chen2021}, nematic order~\cite{Nie2022}, etc. 
While, in (K,Rb)V$_3$Sb$_5$, the evolution of nodal to nodeless SC can be tuned by external pressure~\cite{Guguchia2023b},  
growing evidence reveals nodeless SC with multiple gaps in CsV$_3$Sb$_5$~\cite{Yin2021,Mu2021,Duan2021, Xu2021,Shan2022,Roppongi2023,Zhong2023}. 
The spin-lattice relaxation rate shows a Hebel-Slichter coherence peak just below $T_c$, indicating that CsV$_3$Sb$_5$ is a conventional superconductor~\cite{Mu2021}, consistent with the absence of spontaneous magnetic fields in the superconducting state~\cite{Gupta2022,Shan2022}. Moreover, the SC order parameter of the $A$V$_3$Sb$_5$ family is most likely a spin singlet at ambient pressure~\cite{Mu2021}, reflected in a distinct drop of the Knight shift below $T_c$.  
Although the CDW order is absent in the isostructural Ti-based $A$Ti$_3$Bi$_5$ kagome family~\cite{Liu2023,Yang2024,Yang2023b,Li2023,Rehfuss2024,Yi2023}, the latter exhibits exotic properties similar to the $A$V$_3$Sb$_5$ family~\cite{Yang2024,Li2023}. In addition, 
at low temperatures, CsTi$_3$Bi$_5$ has been reported to be a superconductor. However, whether its SC is intrinsic or arises from impurity phases is still under debate~\cite{Yang2024,Yang2022}. 
The recent discovery of pressure-induced SC and low-energy magnetic excitations in the Cr-based kagome antiferromagnet CsCr$_3$Sb$_5$~\cite{Liu2024b,Wang2025b,Li2025,Wu2025}, whose phase diagram resembles that of iron-pnictide or cuprate superconductors, has further stimulated research on SC in kagome materials.

In kagome superconductors, the interplay between band topology and unconventional superconducting pairing, as well as the impact of spin fluctuations on superconductivity
are far from being fully understood and clearly require further investigation.
Due to the heavy iridium atoms, 
the $M$Ir$_2$ kagome family (with $M$ = Ca, Sr, Ba, Sc, Zr, Th), which belongs to the  well-known Laves phase~\cite{Koshinuma2022,Haldo2015,Horie2020,Xiao2021,Yang2023,Geballe1965}, represents a
suitable system  to investigate the unconventional SC, with $T_c$ values ranging from 2 to 6\,K. 
Unlike the other members, a SC with $T_c \approx$ 2\,K was reported decades ago in ScIr$_2$~\cite{Geballe1965}, and its electronic band structures are available~\cite{Chowdhury2019,Uzunok2020}. However, its superconducting properties have only recently been studied by electrical resistivity and specific heat~\cite{zhu2024anomalous}.
Interestingly, in the ScIr$_{2-x}$Si$_x$ ($0 \le x \le 0.35$) family,
the Ir/Si substitution results in a two-dome-like superconducting phase~(Figure~\ref{fig:phase}a), suggested to be unconventional by the electronic specific-heat data~\cite{zhu2024anomalous}. 

Here, we combine muon-spin spectroscopy ({\textmu}SR), single-crystal x-ray diffraction (SXRD), electrical transport, and magnetization measurements with theoretical band-structure calculations to identify ScIr$_{2-x}$Si$_x$ compounds as a unique family of kagome superconductors. Their superconducting pairing is well described by a two-gap model, where one of the gaps has nodes, indicating an unconventional SC nature. In addition, a structural phase transition, from a high-$T$ cubic phase (centrosymmetric)- to a low-$T$ rhombohedral phase (noncentrosymmetric) is firstly confirmed by our SXRD measurements 
in the parent ScIr$_2$ compound. The low-$T$ phase not only hosts a kagome layer formed by Ir atoms, but it also exhibits a chiral chain of Ir atoms. Therefore, ScIr$_2$ can be classified as a topological chiral crystal that admits a unique band topology, implying fermionic excitations with large Chern numbers, giant helicoid Fermi arcs, etc.

\section{Results and Discussion}
\subsection{Structural Phase Transition}
Since the ScIr$_{2-x}$Si$_x$ family exhibits a two-dome-like superconducting phase diagram with 
maxima located at $x$ = 0 and 0.25 (Figure~\ref{fig:phase}a), we synthesized the corresponding ScIr$_2$ and ScIr$_{1.75}$Si$_{0.25}$ compounds and systematically investigated their superconducting properties. 
\tcr{For ScIr$_{1.75}$Si$_{0.25}$, we found that the actual Si concentration is slightly lower than the nominal value (see below). Hereafter, the actual Si concentration ScIr$_{1.82}$Si$_{0.18}$ will be used instead of ScIr$_{1.75}$Si$_{0.25}$.}
Room-temperature powder x-ray diffraction (XRD) refinements confirm their cubic crystal structure with a space group $Fd\bar{3}m$ (No.~227) (Figure~S1, Supporting Information). 
Each compound exhibits a distinct temperature-dependent electrical resistivity $\rho(T)$ (Figure~\ref{fig:phase}b).
While ScIr$_2$ shows a strongly temperature-dependent $\rho(T)$, ScIr$_{1.82}$Si$_{0.18}$ exhibits an almost temperature independent resistivity in its normal state. At low temperatures, both compounds undergo a superconducting transition at $T_c$ = 1.6 and 3.2\,K, respectively, where the resistivity drops to zero. Interestingly, 
at $T_\mathrm{S} \approx 190$\,K, ScIr$_2$ exhibits a clear anomaly in $\rho(T)$, which is easily visible in the derivative of resistivity with respect to temperature (indicated by an arrow in Figure~\ref{fig:phase}b). 
This anomaly is less evident in the Si-doped ScIr$_2$~\cite{zhu2024anomalous}, and is unlikely to be of magnetic origin, as confirmed by our {\textmu}SR measurements (see below). To investigate the origin of this anomaly, single-crystal x-ray diffraction measurements were performed on both the ScIr$_2$ and ScIr$_{1.82}$Si$_{0.18}$ compounds (see the Experimental Section for details). SXRD refinements of ScIr$_2$ at $T = 80$\,K, i.e., well below $T_\mathrm{S}$, show a significantly better agreement with the rhombohedral space group $R32$ (No.~155) (Figure~\ref{fig:phase}c) than with the cubic group $Fd\bar{3}m$ (No.~227) (Figure~\ref{fig:phase}d). 
Conversely, the SXRD data of ScIr$_{1.82}$Si$_{0.18}$ at both 80\,K and 280\,K are consistent with the $Fd\bar{3}m$ space group (Figure~\ref{fig:phase}e).
The low-$T$ rhombohedral phase is a distorted variant of the high-$T$ cubic phase, reflecting the displacement of Ir atoms from the ideal planar kagome layer. The details of SXRD refinements are summarized in Table~S1 of the Supporting Information.
SXRD refinements suggest that the actual Si concentration is slightly lower than the nominal value, namely, ScIr$_{1.82}$Si$_{0.18}$, consistent with powder XRD results (Figure~S1, Supporting Information). Our SXRD measurements reveal a new structural phase transition
in ScIr$_2$, from the high-$T$ cubic phase (centrosymmetric) to the low-$T$ rhombohedral phase (noncentrosymmetric). 
Note that this structural phase transition can also be viewed as a nematic transition that breaks the rotational symmetry.
Similar to ScIr$_2$, the isostructural PbAu$_2$ superconductor has been
reported to undergo a structural phase transition, this time from
the high-$T$ cubic- to the low-$T$ orthorhombic phase at $T_\mathrm{S}$ $\approx$ 50\,K~\cite{Schoop2015,Xing2016}.
Since the structural phase transition at $T_\mathrm{S}$ in ScIr$_2$ is highly sensitive
to disorder, it is quickly suppressed by Si-doping. 
As the temperature decreases below $T_\mathrm{S}$, the Ir kagome layer remains, albeit slightly distorted (Figure~\ref{fig:phase}g and Table~S1, Supporting Information).
Interestingly, the Ir atoms at 9$d$ sites also form chiral chains in the low-$T$ phase. 
Therefore, ScIr$_2$ can be classified as a topological chiral crystal, 
enabling the formation of unique topological electronic states. 

\subsection{Characterization of Superconductivity} 
The SC of ScIr$_{2-x}$Si$_{x}$ was characterized via measurements of magnetic susceptibility and electrical resistivity. The temperature-dependent magnetic susceptibility $\chi(T)$ in Figure~\ref{fig:SC}a shows clear superconducting transitions at $T_c = 1.8$\,K and 3.3\,K for ScIr$_2$ and ScIr$_{1.82}$Si$_{0.18}$, respectively.
The well-separated ZFC- and FC-curves confirm a type-II SC in both compounds, further supported by our {\textmu}SR measurements.
Both ScIr$_2$ and ScIr$_{1.82}$Si$_{0.18}$ exhibit bulk SC with a diamagnetic screening over 100\% and 70\%, respectively, consistent with previous studies~\cite{zhu2024anomalous}. The zero-field electrical resistivity drops to zero at $T_c$ $\approx$ 1.6\,K and 3.3\,K for ScIr$_2$ and ScIr$_{1.82}$Si$_{0.18}$, respectively.
The upper critical fields determined from the electrical resistivity measured under various magnetic fields (see inset in Figure~\ref{fig:SC}b) are $\mu_0$$H_\mathrm{c2}$(0) = 0.65(3) and 4.8(1)\,T for ScIr$_2$ and ScIr$_{1.82}$Si$_{0.18}$, respectively. 
For both compounds, the upper critical fields are below the Pauli-limiting field (i.e., $\mu_0H_\mathrm{P}$ = 1.86$T_c$). Consequently, the orbital pair-breaking mechanism is dominant in the ScIr$_{2-x}$Si$_x$ family. 

In the Ginzburg-Landau (GL) theory of SC, the coherence length 
$\xi$ can be calculated from $\xi$ =  $\sqrt{\Phi_0/2\pi\,H_{c2}}$, where 
$\Phi_0 = 2.07 \times 10^{3}$\,T~nm$^{2}$ is the quantum of
magnetic flux. With $\mu_{0}H_{c2}(0)$ = 0.65 and 4.8\,T, the calculated $\xi(0)$ values
are 22.5 and 8.3\,nm for ScIr$_2$ and ScIr$_{1.82}$Si$_{0.18}$, respectively. 
The magnetic penetration depth $\lambda_0$ is related to 
the coherence length $\xi$ and the lower critical field $H_{c1}$ 
via $\mu_{0}H_{c1} = (\Phi_0 /4 \pi \lambda^2)[$ln$(\kappa)+ 0.5]$, where 
$\kappa$ = $\lambda_0$/$\xi$ is the GL parameter~\cite{Brandt2003}.
By using $\lambda_0$ = 425(3) and 680(3)\,nm, we find $\mu_{0}H_{c1}(0)$ = 3.14(4) and 1.75(1)\,mT for ScIr$_2$ and ScIr$_{1.82}$Si$_{0.18}$, respectively.

\subsection{TF-{\textmu}SR and Superconducting Pairing} 
The symmetry of superconducting pairings is crucial to understand the mechanism of SC. 
To investigate the superconducting pairing of ScIr$_{2-x}$Si$_{x}$, its temperature-dependent magnetic penetration depth was measured via transverse-field (TF-) {\textmu}SR.
Since the superfluid density $\rho_\mathrm{sc}$ is proportional to the inverse
square of the magnetic penetration depth $\lambda_\mathrm{eff}$, i.e.,  $\rho_\mathrm{sc}$  $\propto$ $\lambda_\mathrm{eff}^{-2}$, the temperature-dependent $\lambda_\mathrm{eff}(T)$ reflects the nature of superconducting pairing (see the Experimental Section for details).
As shown in Figure~\ref{fig:TF-μSR}a,b, the TF-{\textmu}SR measurements were performed in an applied field of 30\,mT for both ScIr$_2$ and ScIr$_{1.82}$Si$_{0.18}$, covering 
the normal and superconducting states.
The enhanced muon-spin relaxation in the superconducting state is clearly visible and is attributed to the formation of a flux-line lattice (FLL) in the mixed state during the field-cooling process, which leads to inhomogeneous field distributions~\cite{Yaouanc2011,Amato2024,Blundell2021}.
By contrast, the weak and temperature-independent relaxation observed in the normal state is attributed to nuclear magnetic moments. 
In the superconducting state, ScIr$_2$ exhibits much larger muon-spin relaxation rates than that of ScIr$_{1.82}$Si$_{0.18}$, indicating shorter magnetic penetration depths in the former case. 
The broadening of the field distribution in the superconducting state is clearly evident in the fast Fourier transform (FFT) of the corresponding TF-{\textmu}SR spectra (see Figure~\ref{fig:TF-μSR}c,d).

To describe the field distribution, TF-{\textmu}SR spectra can be modeled using Eq.~\eqref{eq:TF}. For both compounds, a single oscillation reproduces the experimental data quite well (solid lines in Figure~\ref{fig:TF-μSR}a-d).  
For $T >$ $T_c$, the relaxation rate is small and temperature-independent, while for $T <$ $T_c$, it starts to increase due to the formation of a FLL and the increased superfluid density. At the same time, a diamagnetic field shift $\Delta$$B_\mathrm{dia}$ appears at the onset of the superconducting transition (see the insets in Figure~\ref{fig:TF-μSR}e,f).
The effective magnetic penetration depth, and thus the superfluid density, were calculated from the measured superconducting {\textmu}SR relaxation rates (see the Experimental Section for details).
The inverse square of the effective magnetic penetration depth $\lambda_\mathrm{eff}^{-2}(T)$ versus the reduced
temperature $T$/$T_c$ for ScIr$_2$ and ScIr$_{1.82}$Si$_{0.18}$ is shown in Figure~\ref{fig:TF-μSR}e,f.
Despite having different crystal structures at low temperatures (rhombohedral vs.\ cubic),
the temperature evolution of $\lambda_\mathrm{eff}^{-2}(T)$ for both compounds is quite similar.  

The lines through the data in Figure~\ref{fig:TF-μSR}e,f are fits to Eq.~\eqref{eq:rho} using different models.
Six different models were used to analyze the superfluid density,
including single-gap $s$-, $p$-, $d$-wave, and two-gap ($s+s$)-, ($s+p$)-,
($s+d$)-wave. The derived fitting parameters are summarized in Table~\ref{tab:parameters}.
For both compounds, the fairly weak temperature-dependent superfluid density at low temperatures clearly rules out a $d$-wave model with line nodes (see the blue lines in Figure~\ref{fig:TF-μSR}e,f). The $p$-wave model with point nodes also shows a poor agreement with the experimental data for $0.2 < T/T_c < 0.6$ (see the red lines in the lower insets). The $s$-wave model without gap nodes shows better agreement with the experimental data, 
here reflected in the relatively small $\chi_\mathrm{r}^2$ values (see Table~\ref{tab:parameters}). For the two-gap scenario, the so-called $\alpha$-model was considered. In this case, the superfluid density can be described by $\rho_\mathrm{sc}(T) = w \rho_\mathrm{sc}^{\Delta_1}(T) + (1-w) \rho_\mathrm{sc}^{\Delta_2}(T)$, where $\rho_\mathrm{sc}^{\Delta_1}$ and $\rho_\mathrm{sc}^{\Delta_2}$ are the superfluid densities related to the first ($\Delta_1$) and the second ($\Delta_2$) gaps, and $w$ is a relative weight. For each gap, $\rho_\mathrm{sc}(T)$ is also given by Eq.~\eqref{eq:rho}. All three two-gap models reproduce the experimental data quite well, and yield relatively small $\chi_\mathrm{r}^2$ values. However, the ($s+p$)-wave model is inconsistent with the preserved time-reversal symmetry (TRS) in the superconducting state
of ScIr$_{2-x}$Si$_x$ (see below). For the ($s+d$)-wave model, the estimated weight of the line-nodal gap
(i.e., the $d$-wave component) is about 20\%. Consequently, the superfluid density is still dominated by the nodeless gap (i.e., the $s$-wave component) in both compounds. The ($s+s$)-wave model shows an almost identical goodness of fit to the ($s+d$)-wave model. In general, TF-{\textmu}SR measurements in a higher magnetic field can distinguish between these two cases.
An increased magnetic field suppresses the $s$-type gap, while the $d$-type gap is more robust, making the nodal features more evident in the superfluid density. This approach has previously been used to reveal the unconventional pairing in the CuIr$_2$Te$_4$ superconductor~\cite{Shang2022}. For ScIr$_2$, the ($s+s$)-wave model yields a smaller $\chi_\mathrm{r}^2$ than the $s$-wave model, while both models produce comparable $\chi_\mathrm{r}^2$ values in ScIr$_{1.82}$Si$_{0.18}$ (see Table~\ref{tab:parameters}).

\subsection{Electronic Specific Heat and Superconducting Pairing}

Some of the theoretical models produce virtually indistinguishable
results when analyzing the superfluid density, as reflected by
their comparable $\chi_\mathrm{r}^2$ values. To further validate the
superconducting pairing of ScIr$_{2-x}$Si$_x$, the zero-field
electronic specific heat $C_\mathrm{e}/T$ 
was analyzed using the aforementioned models (see the Experimental Section for details). As shown in Figure~\ref{fig_Cp}, the lines through the data are fits to Eq.~\eqref{eq:entropy} using different models. Clearly, the single-gap $d$-wave model deviates significantly from the data for both compounds. The $s$-wave model reproduces the $C_\mathrm{e}/T$ data reasonably well for $T/T_c > 0.4$, but it deviates from them at lower temperatures, as reflected by the large $\chi_\mathrm{r}^2$ values (see Table~\ref{tab:parameters}). Similar to the TF-{\textmu}SR results, the $p$-wave model shows reasonable agreement with the experimental data, but it is inconsistent with the preserved TRS. In the case of two-gap models, the electronic specific heat can be modeled by $C_\mathrm{e}(T)/T = wC_\mathrm{e}^{\Delta_1}(T)/T + (1-w)C_\mathrm{e}^{\Delta_2}(T)/T$~\cite{Bouquet2001}. 
Here, each term represents the contribution to the specific heat
of the individual gaps, with $w$, $\Delta_1$, and $\Delta_2$ being the
same parameters as for the superfluid-density fits in Figure~\ref{fig:TF-μSR}e,f.
The two-gap ($s+s$) and ($s+p$)-wave models both show a poor
agreement with the $C_\mathrm{e}/T$ data at $T/T_c < 0.3$, yielding large $\chi_\mathrm{r}^2$ values. 
Conversely, the ($s+d$)-wave model exhibits a much better agreement with the $C_\mathrm{e}/T$ data across the full temperature range, resulting in the smallest deviation (see Table~\ref{tab:parameters}). The ($s+d$)-wave pairing is also consistent with the nonlinear field-dependent electronic specific heat coefficient in the superconducting state~\cite{zhu2024anomalous}.
Figure~S2 in the Supporting Information presents the separate $s$ and $d$
components of the superfluid density and electronic specific heat.
By combining the temperature-dependent superfluid density and electronic specific heat, both of which are well described by a two-gap ($s+d$)-wave model, we provide solid evidence for unconventional SC in the ScIr$_{2-x}$Si$_x$ family. In addition, the multigap characteristics are further supported by band-structure calculations (see below).

\subsection{ZF-{\textmu}SR and Preserved TRS} 
Since some of the kagome superconductors are known to break TRS in either
the charge-ordered or the superconducting state (see Table~S2 in the Supporting Information),
zero-field (ZF-) {\textmu}SR measurements were performed in the normal-
and superconducting states of the ScIr$_{2-x}$Si$_x$ family to verify
a possible TRS breaking. This technique is highly sensitive to the weak spontaneous fields
that are expected to arise in this case~\cite{Yaouanc2011,Amato2024,Blundell2021}.
As can be seen
in Figure~\ref{fig:ZF-muSR}, no coherent oscillations or fast decays were identified in
the ZF-{\textmu}SR spectra of ScIr$_2$ and ScIr$_{1.82}$Si$_{0.18}$, indicating the absence of 
magnetic order or fluctuations. Here, the muon-spin relaxation is mainly determined by the randomly oriented nuclear moments in the absence of an external magnetic field. Therefore, ZF-{\textmu}SR spectra can be modeled using a combined Gaussian- and
Lorentzian Kubo-Toyabe relaxation function. The solid lines through the data in Figure~\ref{fig:ZF-muSR} are fits to Eq.~\eqref{eq:ZF}, yielding muon-spin relaxation rates
$\sigma_\mathrm{ZF}$ = 0.241(3) and $\Lambda_\mathrm{ZF}$ = 0.025(4)\,{\textmu}s$^{-1}$, and $\sigma_\mathrm{ZF}$ = 0.254(3) and  $\Lambda_\mathrm{ZF}$ = 0.031(4)\,{\textmu}s$^{-1}$ at $T$ = 0.3 and 4\,K for ScIr$_2$, respectively. For ScIr$_{1.82}$Si$_{0.18}$, we obtain
$\sigma_\mathrm{ZF}$ = 0.221(3) and $\Lambda_\mathrm{ZF}$ = 0.011(3)\,{\textmu}s$^{-1}$, and $\sigma_\mathrm{ZF}$ = 0.217(3) and $\Lambda_\mathrm{ZF}$ = 0.010(3)\,{\textmu}s$^{-1}$ at $T$ = 0.3 and 4\,K, respectively.
For both compounds, the relaxation rates in the normal- and superconducting states are almost identical, visually
confirmed by the overlapping ZF-{\textmu}SR spectra. The absence of an additional {\textmu}SR relaxation below $T_c$ rules out the possibility of TRS breaking in the superconducting state of ScIr$_{2-x}$Si$_x$. In turn, the preservation of TRS excludes a chiral $p$- or ($s+p$)-wave pairing in the ScIr$_{2-x}$Si$_x$ kagome superconductors. This differs significantly from the CaPtAs and LaPt$_3$P superconductors, where the ($s+p$)-wave model or the chiral $d$-wave model have been 
proposed to explain the gap nodes and the broken TRS in their superconducting state~\cite{Shang2020,Biswas2021}.

\subsection{Electronic Band Structures and Topology Aspects} 
To gain further insight into the electronic and superconducting properties
of the ScIr$_{2-x}$Si$_x$ family, we also performed band-structure calculations 
using the density-functional theory (DFT). In the parent compound ScIr$_2$, both SXRD and electrical resistivity
measurements reveal a structural phase transition from the high-$T$ cubic phase
to the low-$T$ rhombohedral phase, which is suppressed in the Si-doped ScIr$_2$.
The electronic band structures of both phases were calculated and
the respective Brillouin zones are shown in Figure~\ref{fig_DFT}a,b. 
The band structures and density of states (DOS) of the cubic phase of ScIr$_2$ are shown in Figure~\ref{fig_DFT}c,d, while 
the analogous results for ScIr$_{1.82}$Si$_{0.18}$ are shown in Figure~\ref{fig_DFT}e,f, the stoichiometry of the latter being determined by SXRD measurements (Table~S1, Supporting Information).
The corresponding results for the rhombohedral phase of ScIr$_2$
are shown in Figure~\ref{fig_DFT}g,h. 
The band structures of ScIr$_{1.75}$Si$_{0.25}$
were also calculated and are presented in Figure~S3 of the Supporting
Information, showing similar results. 

In both phases, the Ir-5$d$ and Sc-4$d$ orbitals dominate the DOS
close to the Fermi level $E_\mathrm{F}$. At the same time, the
contribution from the Si-3$p$ orbitals is negligible in the Si-doped compounds.
The dominance of the Ir 5$d$-orbitals suggests a large SOC
effect, which plays an important role in determining the electronic
properties of ScIr$_{2-x}$Si$_x$. Band splitting due to the SOC is
significant along the $\Gamma-L$ direction of ScIr$_2$, where
also the flat bands exist, the latter being pushed closer to $E_\mathrm{F}$ by the SOC.
At the $L$ point of ScIr$_2$, the estimated band splitting $E_\mathrm{SOC}$ is
about 275\,meV in the cubic phase (Figure~\ref{fig_DFT}c) and 264\,meV in the rhombohedral phase (Figure~\ref{fig_DFT}g). 
The band structures calculated by ignoring the SOC are presented in Figures~S4 in the Supporting Information.
Apart from a shift in the Fermi level towards lower energies, the electronic bands of the cubic ScIr$_{1.82}$Si$_{0.18}$ are almost identical to those of ScIr$_2$, with estimated DOS values at the $E_\mathrm{F}$ of 2.66 and 5.19\,states/(eV f.u.), respectively. 
While for the rhombohedral ScIr$_2$, the DOS is about 2.65\,states/(eV f.u.). 
Furthermore, up to three and two different electronic bands crossing the Fermi level
were identified in the rhombohedral and cubic structures, respectively, thus strongly corroborating the observed
multigap SC (see Figures~S5 and S6 in the Supporting Information).

In the cubic phase, Ir atoms form a kagome layer along the [111]-direction (Figure~\ref{fig:phase}f).
After undergoing a structural phase transition to the rhombohedral phase at
$T_\mathrm{S} \approx 190$\,K (see Figure~\ref{fig:phase}b), the Ir kagome
layer is preserved, but it is now found in the [001]-direction (Figure~\ref{fig:phase}g). 
As shown in Figure~\ref{fig_DFT}c, due to the large SOC of Ir-5$d$ orbitals,
flat bands along the $\Gamma-L$ direction can be clearly identified in
the cubic ScIr$_2$. Interestingly, one of these flat bands is located very
close to the Fermi level (just 10\,meV below $E_\mathrm{F}$), resulting
in a sharp DOS peak near $E_\mathrm{F}$, with a value of about 5.19\,states/(eV f.u.).
The rhombohedral phase of 
ScIr$_2$ lacks an inversion center and its antisymmetric SOC lifts the
degeneracy of the electronic bands. As a consequence, each flat
band splits into two bands along the $\Gamma-L$ direction
(Figure~\ref{fig_DFT}g), and the DOS near $E_\mathrm{F}$ is
significantly suppressed to about 2.65\,states/(eV f.u.).
Upon substituting Ir by Si atoms, the flat bands along the $\Gamma-L$
direction are shifted upwards by about 300\,meV in ScIr$_{1.82}$Si$_{0.18}$ (see Figure~\ref{fig_DFT}e),
indicating that the silicon substitution introduces a large amount of  hole carriers into the ScIr$_{2-x}$Si$_{x}$ family. Since the structural phase transition is suppressed upon Si doping,
we propose that the flat bands and the DOS peaks are responsible
for the structural instability in the parent cubic ScIr$_2$, driving its structural transition at $T_\mathrm{S} \approx 190$\,K.
In the ScIr$_{2-x}$Si$_{x}$ family, electrons on the flat bands
exhibit strong correlations and large effective masses.
Since flat bands account for most of the DOS at $E_\mathrm{F}$, this may explain the unconventional superconducting pairing revealed by both TF-{\textmu}SR (Figure~\ref{fig:TF-μSR}) and specific-heat measurements (Figure~\ref{fig_Cp}) in the ScIr$_{2-x}$Si$_{x}$ family.

The $X$ point of the cubic phase hosts Dirac fermions with four-fold
degeneracy~\cite{yu_encyclopedia_2022} and, in the ScIr$_2$ parent compound,
it is located about 369\,meV below $E_\mathrm{F}$ (see Figure~\ref{fig_DFT}c).
Upon Si doping, such a band crossing 
shifts up to reach almost the Fermi level (Figure~\ref{fig_DFT}e).  
Hence, the ScIr$_{2-x}$Si$_x$ family represents one of the few examples,
where topological states can be significantly tuned by chemical substitution.
In the rhombohedral phase, due to the absence of inversion symmetry, the Dirac point
splits into two Weyl points with topological charges $C = \pm 1$.
Interestingly, the space group $R32$ (No.~155) of the low-$T$ rhombohedral phase
belongs to one of the chiral space groups. 
As shown by arrows in Figure~\ref{fig:fermi_arc}a, the Ir atoms on 9$d$ sites (see Table~S1, Supporting Information) form a chiral chain 
along the [001]-direction. Therefore, ScIr$_2$ can be classified as a topological
chiral crystal, which exhibits a plethora of exotic quantum phenomena
driven by a combination of structural and electronic chirality~\cite{Yan2024,Hasan2021,Narang2021,Bradlyn2016,Chang2018,Fernando2017,Tan2022}. 
The band crossings at the $\Gamma$ and $Z$ points host Weyl fermions with topological charges of $C = 3$~\cite{yu_encyclopedia_2022}.  
As shown in Figure~\ref{fig:fermi_arc}b, the (001) surface states of ScIr$_2$ appear at the $\bar{\Gamma}$ point. The giant helicoid Fermi arcs that span the entire surface Brillouin zone are a clear hallmark
of topological chiral crystals~\cite{sanchez_topological_2019,Rao2019}. 
Owing to the above facts, ScIr$_2$ represents an ideal platform to explore the physics of chiral topological materials.

\section{Discussion} 
Upon substituting Ir by Si, ScIr$_{2-x}$Si$_x$ ($0 \le x \le 0.35$) exhibits a two-dome-like superconducting phase (Figure~\ref{fig:phase}a), with a maximum $T_c$ reaching 3.3\,K for $x = 0.25$. According to previous work~\cite{zhu2024anomalous}, in ScIr$_2$, the flat band along the $\Gamma-L$ direction is slightly off the Fermi level, resulting in a decrease of DOS at $E_\mathrm{F}$ in the first SC dome. As the Si content is increased further to $x = 0.25$, the flat band moves closer to $E_\mathrm{F}$, resulting in an increased DOS in the second SC dome. 
At the same time, our SXRD and electrical resistivity measurements reveal that the parent ScIr$_2$ compound undergoes a structural phase transition (from the cubic- to the rhombohedral phase) at $T_\mathrm{S} \approx 190$\,K, absent in ScIr$_{1.82}$Si$_{0.18}$. 
\tcr{Since the Si doping suppresses the structural phase transition at $x \gtrsim$ 0.15 (see Figures~S7 and S8, Supporting Information)},  
most likely,  
in the first SC dome, ScIr$_{2-x}$Si$_x$ adopts a rhombohedral crystal structure. Upon further increasing the Si content, in the second SC dome, ScIr$_{2-x}$Si$_x$ adopts a cubic crystal structure.
Considering that previously the band structures of ScIr$_{2-x}$Si$_x$ were calculated based on a cubic structure~\cite{zhu2024anomalous}, the evolution of the flat band with Si doping might not properly explain the changes of DOS and $T_c$ values. 
This is clearly reflected by our band-structure calculations based on the rhombohedral structure.
As the low-$T$ rhombohedral phase of ScIr$_2$ lacks an inversion center, its antisymmetric SOC lifts the degeneracy of electronic bands. As a consequence, each flat band in the cubic phase (see Figure~\ref{fig_DFT}c) splits into two bands along the $\Gamma-L$ direction (Figure~\ref{fig_DFT}g), resulting in a significantly reduced DOS near $E_\mathrm{F}$, which explains the lower $T_c$ values in the first SC dome. \tcr{To fully characterize the low-$T$ crystal structure of the ScIr$_{2-x}$Si$_x$ family, particularly for $x \ge 0.1$, x-ray diffraction measurements using a synchrotron facility are highly desirable.}

From the combined temperature-dependent superfluid density and electronic specific heat, the SC of ScIr$_{2-x}$Si$_x$ is mostly consistent with a two-gap $(s+d)$-wave model (see details in Table~\ref{tab:parameters}). A single-gap $d$-wave model proposed previously deviates significantly from the experimental data~\cite{zhu2024anomalous}. The multiband feature is strongly supported by the band-structure calculations (Figure~\ref{fig_DFT} and  Figures~S5 and S6, Supporting Information). The ($s+d$)-wave model yields a zero-temperature effective magnetic penetration depth $\lambda_0 = 425$ and 680\,nm for ScIr$_2$ and ScIr$_{1.82}$Si$_{0.18}$, respectively. In the $T_c$ vs.\ $\lambda^{-2}_\mathrm{eff}$ phase diagram (see Figure~S9, Supporting Information), both the compounds studied here are clearly away from the area of conventional superconductors, e.g., Nb, Pb. The parent compound ScIr$_2$ follows the trend observed in other kagome superconductors. Despite the different $T_c$ values, the CeRu$_2$, LaRu$_3$Si$_2$, Ta$_2$V$_{3.1}$Si$_{0.9}$, and Re$_2$(Hf,Zr) kagome compounds show nodeless SC~\cite{Mielke2021,Mielke2022,Graham2024,Mandal2022,Mandal2025}, with the CeRu$_2$ being isostructural to the high-$T$ cubic phase of ScIr$_{2-x}$Si$_x$ (Table S2, Supporting Information). In particular, ScIr$_2$ is very close to the $A$V$_3$Sb$_5$ family, where multigap SC has also been reported by {\textmu}SR and scanning tunneling microscopy techniques~\cite{Mielke2022b,Guguchia2023b,Gupta2022,Gupta2022b,Xu2021}.  
The $\lambda_\mathrm{eff}^{-2}(T)$ of (K,Rb)V$_3$Sb$_5$ is almost linear in temperature at $T <$ 0.3\,K, indicating the presence of gap nodes, too~\cite{Guguchia2023}. Similar to the $(s+d)$-wave model of the ScIr$_{2-x}$Si$_x$ family (see Figures~\ref{fig:TF-μSR} and \ref{fig_Cp}), the $\lambda_\mathrm{eff}^{-2}(T)$ of (K,Rb)V$_3$Sb$_5$ is also well described by a two-gap model, where one of the gaps has nodes and the other
does not. In addition, upon applying external pressure, a crossover from nodal- to nodeless SC is observed in (K,Rb)V$_3$Sb$_5$~\cite{Guguchia2023b}. 
Clearly, it would be interesting to investigate the pressure effect also on the ScIr$_{2-x}$Si$_x$ family. 
\tcr{In addition, a multigap model incorporating both an anisotropic and an isotropic $s$-wave gap on two distinct Fermi surfaces has been proposed to describe the multiband SC observed in CsV$_3$Sb$_5$~\cite{Roppongi2023}.}
Finally, Si-doping increases the $T_c$ up to 3.3\,K in ScIr$_{1.82}$Si$_{0.18}$, positioning it in the band of cuprates in the $T_c$ vs.\ $\lambda^{-2}_\mathrm{eff}$ phase diagram, further suggesting the unconventional nature of SC of the ScIr$_{2-x}$Si$_x$ kagome family. 

\tcr{In a multiband system, the microscopic mechanism behind the ($s+d$)-wave pairing remains an open problem. Most likely such a pairing arises from a combination of electron-electron and electron-phonon interactions. In kagome superconductors, two possible pairing mechanisms have been considered: one due to conventional electron-phonon coupling and one arising from spin- or charge fluctuations driven by electron correlation. The former usually results in an $s$-wave pairing, while the latter can lead to various unconventional pairing symmetries (e.g., $p$-, $d$-, or $f$-wave)~\cite{Wu2021,Wen2022,Romer2022}. Of the various electronic bands that cross the Fermi level in the ScIr$_{2-x}$Si$_x$ family, only one band (marked in red in Figures~S5 and S6, Supporting Information) exhibits distinct flat-band features along the $\Gamma-L$ direction, while these features are less pronounced in the other bands (see the blue and green bands in Figures~S5 and S6, Supporting Information). Therefore, as with the (K,Rb,Cs)V$_3$Sb$_5$ family~\cite{Guguchia2023b,Roppongi2023} and given the presence of VHSs near the $E_\mathrm{F}$~\cite{zhu2024anomalous}, the enhanced electron correlation in the red band leads to an unconventional pairing mechanism, e.g., via spin- or charge fluctuations, while the conventional electron-phonon coupling mechanism dominates in the other bands. Therefore, different pairing mechanisms might coexist in the ScIr$_{2-x}$Si$_x$ family. 
}

In the low-$T$ rhombohedral phase, ScIr$_2$ not only hosts a kagome layer formed by Ir atoms, but it also hosts 
an Ir chiral chain.
Therefore, ScIr$_2$ can be classified as a unique topological chiral crystal that exhibits unconventional SC. 
\tcr{Since the chiral structure breaks the inversion symmetry, ScIr$_2$ also belongs to the non-centrosymmetric superconductor (NCSC) family. In the NCSCs, the lack of inversion symmetry gives rise to an antisymmetric SOC, which leads to spin-split Fermi surfaces. Consequently, their pairing states are not constrained to be purely singlet or triplet, and mixed-parity pairing may occur~\cite{Bauer2012,Sungkit2014,Smidman2017}. The strength of SOC generally
determines the degree of mixing of singlet- and triplet states. 
According to the band-structure calculations (see Figure~\ref{fig_DFT}), the estimated band splitting due to the antisymmetric SOC is about $E_\mathrm{SOC} \approx 260$\,meV at the $L$ point. This gives $E_\mathrm{SOC}/k_\mathrm{B} T_c \approx 2000$, which is comparable
to that of the Li$_2$Pt$_3$B, CePt$_3$Si, and CaPtAs NCSCs, all of which exhibit nodal SC~\cite{Smidman2017,yuan2006,Bonalde2005,Shang2020}. The presence of a large band splitting, in addition to multigap SC with gap nodes, suggests that ScIr$_2$ is a promising candidate for exploring mixed-parity pairing. However, whether this crystal structure ($R32$, No~155), exhibits a TRS-preserved mixed-parity pairing state compatible with the anisotropic SOC, requires further theoretical investigation.}

Resembling the other topological chiral crystals, ScIr$_2$ also shows spin-polarized Fermi arc states spanning the entire surface Brillouin zone (Figure~\ref{fig:fermi_arc}b), which can be verified by 
photoemission techniques.
Chiral crystals usually host exotic quantum phenomena driven by the interplay of band topology, SOC, and electronic correlations~\cite{Yan2024,Hasan2021,Narang2021,Bradlyn2016,Chang2018,Fernando2017,Tan2022}.
Although substantial progress has been made in understanding their
topological aspects, their unconventional SC 
remains largely unexplored due to the limited availability of suitable
chiral crystals.  So far, only a few  chiral crystals have been found to exhibit SC~\cite{Carnicom2018,Sun2015,yuan2006,nishiyama2007,Karki2010,Amon2018,Tsvyashchenko2016,Joshi2015,Koizumi2024,Lv2020}, and most of them behave as conventional superconductors~\cite{Smidman2017}. 
\tcr{In chiral crystals, due to the presence of symmetry-protected surface states, such as chiral Fermi arcs or parallel spin-momentum locking, the chiral structure can result in distinct behavior between the bulk- and surface SC. While the bulk SC may exhibit conventional pairing symmetries, the surface states often demonstrate unconventional pairings (e.g., chiral $p$-wave) or topological SC. Significant theoretical work has focused on studying heterostructures consisting of conventional $s$-wave superconductors and topological materials, including topological insulators, Weyl semimetals, Dirac semimetals, and nodal-line semimetals~\cite{Fu2008,Qi2011}. The aim of these studies is to achieve the topological proximity effect by inducing topological SC at the heterostructure interface. However, chiral crystals have rarely been considered. Recently, type-I bulk SC has been observed in the NbGe$_2$ chiral crystal below $T_c = 2$\,K~\cite{Lv2020}, while chiral surface states have been observed on the (100) surface, potentially leading to topological SC, which differs significantly from the pairing symmetry of the bulk~\cite{Yao2025}. Li$_2$Pt$_3$B is known to exhibit spin-triplet pairing in the bulk~\cite{yuan2006}, while it shows a nodal topological SC with dispersionless surface Majorana modes~\cite{Gao2022}. Additionally, topological SC has been theoretically investigated in B20-RhSi and RhGe~\cite{Lee2021,Mardanya2024}, yet, unfortunately, both materials lack bulk SC.
We found that the  ScIr$_{2-x}$Si$_x$ family is a valuable platform for investigating both bulk- and surface SC. In this case, the former is
closely related to the electron correlations in the flat bands, while the latter to the chirality-induced topological band and surface states. In the future, the Kerr rotation technique, low-energy {\textmu}SR, and scanning tunneling microscopy could be used to search for possible
unconventional pairings (e.g., chiral $p$-wave) or topological SC in the surface states of ScIr$_{2-x}$Si$_x$ single crystals.}

Very recently, unconventional SC has been reported in La(Rh,Ir)Si chiral crystals with double helixes~\cite{Shang2025}.
While LaRhSi behaves as a fully-gapped superconductor, LaIrSi shows SOC induced topological nodal-line SC.
Resembling the La(Rh,Ir)Si chiral crystals, $M$Ir$_2$ ($M$ = Ca, Sr, Ba, Sc, Zr, Th) superconducting family~\cite{Koshinuma2022,Haldo2015,Horie2020,Xiao2021,Yang2023}, where SOC can be significantly tuned as well, represents another promising material platform to investigate the SOC effect on the unconventional SC and its interplay with topological bands. Indeed, unconventional superconducting properties have been observed in (Ca,Sr)Ir$_2$ under pressure~\cite{yang2020b,li2021}.

\section{Conclusion}
In summary, our study provides clear evidence for unconventional SC in the newly discovered ScIr$_{2-x}$Si$_x$ family
with a kagome lattice. The temperature-dependent superfluid density and the electronic specific heat both can be accurately described by a two-gap $(s+d)$-wave model, in which one of the gaps has nodes and the other does not. The unconventional SC of ScIr$_{2-x}$Si$_x$ is closely related to the flat bands near the Fermi level. Additionally, ScIr$_2$ undergoes a structural phase transition from a high-$T$ cubic phase to a low-$T$ rhombohedral phase. Since the cubic phase is centrosymmetric, while the rhombohedral phase is noncentrosymmetric, the latter belongs to a chiral space group. Hence, additional exotic topological features are expected to emerge at low temperatures. 
ScIr$_2$ represents a very rare case, where Ir atoms host both a kagome lattice and structural chirality -- both of which being hot topics in condensed matter physics. The ScIr$_{2-x}$Si$_x$ family provides an ideal platform for investigating the interplay between unconventional and topological SC, flat bands, and chirality.

\section{Experimental Section}
\noindent\textbf{Sample Preparation and Characterization}\\
Polycrystalline ScIr$_{2-x}$Si$_x$ ($x$ = 0, 0.1, and 0.25) samples were prepared by arc melting stoichiometric mixtures of Sc pieces (99.9\%, Alfa Aesar), Ir powders (99.9\%, ChemPUR), and Si chunks (99.9999\%, Alfa Aesar) in a high-purity argon atmosphere. 
To improve sample homogeneity, the ingots were flipped and re-melted several times. The crystal structure of ScIr$_{2-x}$Si$_x$ samples was checked via powder x-ray diffraction at room temperature using a Bruker D8 diffractometer with Cu K$\alpha$ radiation ($\lambda$ = 1.5406~\AA{}). 
The electrical-resistivity and magnetic susceptibility measurements were performed on a Quantum Design physical property measurement system and a magnetic property measurement system.

\vspace{5mm}
\noindent\textbf{Single-crystal X-ray Diffraction}\\
Single crystals of ScIr$_{2-x}$Si$_x$ with $\sim$20~{\textmu}m in diameter were sifted out from the crushed arc-melted alloy samples and were mounted on the MiTeGen MicroMounts loop for the X-ray structure determination. Measurements were performed in the temperature range bewteen 80\,K  and 280\,K using the STOE STADIVARI diffractometer equipped with a Dectris EIGER 1M 2R CdTe detector and with an Anton Paar Primux 50 Ag/Mo dual-source using Mo K$\alpha$ radiation ($\lambda$ = 0.71073~\AA{}) from a micro-focus X-ray source and coupled with an Oxford Instruments Cryostream 800 jet. Data reduction was performed with X-Area package, while crystal structure was solved and refined using the JANA2020 software package~\cite{Petricek2020}.

\vspace{5mm}
\noindent\textbf{{\textmu}SR Experiments}\\
The {\textmu}SR experiments were conducted at the Versatile Muon Spectrometer (VMS) on the $\pi$E1 beamline of the Swiss muon source (S{\textmu}S) at the Paul Scherrer Institut (PSI) in Villigen, Switzerland. 
For TF-$\mu$SR measurements, the applied magnetic field (i.e., 30\,mT) was perpendicular to the muon-spin direction, and the samples were cooled in an applied magnetic field down to the base temperature (i.e., 0.3\,K).
For the ZF-$\mu$SR measurements, to exclude the possibility of stray magnetic fields, all the magnets were preliminarily degaussed, and an active field-compensating facility was used. Both the ZF- and TF-{\textmu}SR spectra were collected upon heating the samples.


\vspace{5mm}
\noindent\textbf{Analysis of the {\textmu}SR Spectra}\\
All the {\textmu}SR spectra were analyzed by means of the \texttt{musrfit} software package~\cite{Suter2012}. 
For the TF-{\textmu}SR spectra, the time evolution of the asymmetry was modeled by:
\begin{equation}
\label{eq:TF}
A_{\mathrm{TF}}(t) = A_s \cos(\gamma_\text{\textmu} B_\mathrm{s} t + \phi) \exp \left( -\frac{\sigma^2 t^2}{2} \right)+ A_{\text{bg}} \cos(\gamma_\text{\textmu} B_{\text{bg}} t + \phi).
\end{equation}
Here $A_\mathrm{s}$ and $A_\mathrm{bg}$ are the initial muon-spin asymmetries for muons implanted in the sample and the sample holder, respectively; $B_\mathrm{s}$ and $B_\mathrm{bg}$ are the local fields sensed by implanted muons in the sample and the sample holder (i.e., the copper plate); $\gamma_\text{\textmu}$/2$\pi$ = 135.53~MHz/T is the muon gyromagnetic ratio, $\phi$ is the shared initial phase, and $\sigma$ is a Gaussian relaxation rate reflecting the field distribution inside the sample. In the superconducting state, the derived $\sigma$
includes contributions from both the FLL ($\sigma_\mathrm{sc}$) and the temperature-independent relaxation due to the nuclear moments ($\sigma_\mathrm{n}$). Then, the contribution from the superconducting state can be extracted by subtracting the $\sigma_\mathrm{n}$ contribution quadratically, i.e., $\sigma_\mathrm{sc} = \sqrt{\sigma^2 - \sigma_\mathrm{n}^2}$.
\tcr{For ScIr$_{1.82}$Si$_{0.18}$, the presence of the non-superconducting Ir$_3$Si$_5$ phase has a negligible effect on the temperature evolution of $\sigma_\mathrm{sc}$ due to its extremely weak nuclear muon-spin relaxation rate ($\sigma_\mathrm{n} <$ 0.03 {\textmu}s$^{-1}$).}
Since $\sigma_{\text{sc}}$ is directly related to the effective magnetic penetration depth, and thus to the superfluid density ($\sigma_\mathrm{sc} \propto \lambda_\mathrm{eff}^{-2} \sim \rho_\mathrm{sc}$), the superconducting gap and its symmetry can be investigated by measuring the temperature-dependent $\sigma_\mathrm{sc}$. The effective magnetic penetration depth $\lambda_\mathrm{eff}$ can be calculated from the measured $\sigma_\mathrm{sc}$ according to $\sigma_\mathrm{sc}^2(T)/\gamma_\text{\textmu}^2 = 0.00371 \Phi_0^2/ \lambda_\mathrm{eff}^{4}(T)$~\cite{Barford1988,Brandt2003}, where $\Phi_0$ is the flux quantum.

In the ZF-{\textmu}SR case, the relaxation is mainly determined by the randomly oriented nuclear magnetic moments of ScIr$_{2-x}$Si$_x$ in the absence of external magnetic field. Therefore, their ZF-{\textmu}SR spectra can be modeled using a phenomenological relaxation function, consisting of a combination of Gaussian- and Lorentzian Kubo-Toyabe relaxations~\cite{Kubo1967,Yaouanc2011,Amato2024,Blundell2021}:
\begin{equation}
\label{eq:ZF}
A_\mathrm{ZF}(t) = A_\mathrm{s}\left[\frac{1}{3} + \frac{2}{3}(1 - 
\sigma_\mathrm{ZF}^{2}t^{2} - \Lambda_\mathrm{ZF} t)\,
\mathrm{e}^{(-\frac{\sigma_\mathrm{ZF}^{2}t^{2}}{2}-\Lambda_\mathrm{ZF} t)}\right] + A_\mathrm{bg}.
\end{equation}
%
Here, $A_\mathrm{s}$ and $A_\mathrm{bg}$ are the same as in TF-{\textmu}SR, and $\sigma_\mathrm{ZF}$ and 
$\Lambda_\mathrm{ZF}$ represent the zero-field Gaussian and Lorentzian relaxation rates, respectively.

\vspace{5mm}
\noindent\textbf{Superconducting Gap Symmetry}\\
To extract the superconducting gap and its symmetry of ScIr$_{2-x}$Si$_x$, various models were used to analyze the  temperature-dependent superfluid density $\rho_\mathrm{sc}(T)$ [$\propto \lambda_\mathrm{eff}^{-2}(T)$], generally described by the following relation: 
\begin{equation}
\label{eq:rho}
\rho_\mathrm{sc}(T) = 1 + 2 \left\langle \int_{\Delta_k}^{\infty} \frac{E}{\sqrt{E^2 - \Delta_k^2}} \frac{\partial f}{\partial E} dE \right\rangle_\mathrm{FS}.
\end{equation}
Here, $f = \left( 1 + e^{E/k_BT} \right)^{-1}$ is the Fermi function, $E(\epsilon) = \sqrt{\epsilon^2 + \Delta_k^2(T)}$ 
is the excitation energy of quasiparticles, with $\epsilon$ the electron energies measured relative to the chemical potential (Fermi
energy), and $\left\langle\right\rangle_\mathrm{FS}$ represents an average over the Fermi surface~\cite{tinkham1996,Prozorov2006}; $\Delta_k(T) = \Delta(T) g_k$ is an angle-dependent gap function, where $\Delta$ is the maximum gap value and $g_k$ is the angular dependence of the gap, equal to 1, $\sin\theta$, and $\cos2\phi$ for the $s$-, $p$-, and $d$-wave models, respectively, where $\phi$ and $\theta$ are the azimuthal angles (see details also in Table~\ref{tab:parameters}). The temperature-dependent gap is assumed to follow $\Delta(T) = \Delta_0 \tanh \{1.82 [ 1.018 (T_c/T - 1)]^{0.51}\}$, where $\Delta_0$ is the zero-temperature gap value~\cite{Carrington2003}.

The zero-field specific heat data taken from Ref.~\cite{zhu2024anomalous} were also analyzed to reveal the superconducting pairings.
To subtract the phonon contribution from the specific heat, 
the normal-state specific heat was fitted to the expression $C/T = \gamma_\mathrm{n} + \beta T^2 + \delta T^4$, 
where $\gamma_\mathrm{n}$ is the normal-state electronic specific-heat coefficient, $\beta$ and $\delta$ are the phonon specific-heat coefficients.  From this, we derive $\gamma_\mathrm{n}$ = 14.8(2)\,mJ/mol-K$^2$, $\beta$ = 1.6(4)\,mJ/mol-K$^4$, and $\delta$ = 0.03(2)\,mJ/mol-K$^6$ for ScIr$_2$, while for ScIr$_{1.82}$Si$_{0.18}$, $\gamma_\mathrm{n}$ = 16.6(1)\,mJ/mol-K$^2$, $\beta$ = 0.20(3)\,mJ/mol-K$^4$, and $\delta$ = 0.016(2)\,mJ/mol-K$^6$. After subtracting the phonon contribution ($\beta$$T^2$ + $\delta$$T^4$)
from the specific-heat data, the electronic specific heat divided by $\gamma_\mathrm{n}$, i.e., $C_e/\gamma_\mathrm{n}T$, were obtained, which are presented in Figure~\ref{fig_Cp} versus the reduced temperature $T/T_c$ for both ScIr$_2$ and ScIr$_{1.82}$Si$_{0.18}$.
The contribution of the superconducting phase to entropy can
be calculated following the BCS expression~\cite{tinkham1996}:
\begin{equation}
	\label{eq:entropy}
	S(T) = -\frac{6\gamma_\mathrm{n}}{\pi^2 k_\mathrm{B}} \int^{\infty}_0 [f\mathrm{ln}f+(1-f)\mathrm{ln}(1-f)]\,\mathrm{d}\epsilon,
\end{equation}
where $f$ is the same as in Eq.~\eqref{eq:rho}.
Then, the temperature-dependent electronic specific heat in the superconducting 
state can be calculated from $C_\mathrm{e} =T \frac{dS}{dT}$. 

\vspace{5mm}
\noindent\textbf{Electronic Band-Structure Calculations}\\
First-principles calculations were performed based on the density functional
theory (DFT), as implemented in the Vienna 
\texttt{ab}-initio simulation package
(VASP) package~\cite{Kresse1996kl,Kresse1996vk}. Projector
augmented wave (PAW) pseudo-potentials were adopted in the calculation~\cite{Kresse1999wc,Blochl1994zz}.
The generalized gradient approximation with the Perdew-Burke-Ernzerhof (PBE) realization~\cite{Perdew1996iq} was used for the exchange-correlation functional.
The valence electrons treated in the calculations include Ir ($5d^86s^1$), Sc ($3d^24s^1$), and Si ($3s^23p^2$). The kinetic energy
cutoff was fixed to 400 eV. For the self-consistent calculations, the
Brillouin zone (BZ) integration was performed on a $\Gamma$-centered mesh
of $15 \times 15 \times 15$ $k$-points. The energy convergence criteria
was set to 10$^{-7}$\,eV.
In order to study Fermi surface of ScIr$_{2-x}$Si$_x$, a Wannier tight binding Hamiltonian consisting of Ir-$5d$, Ir-$6s$, Sc-$3d$, Sc-$4s$, Si-$3s$ and Si-$3p$ orbitals were constructed using the Wannier90 package~\cite{pizzi_wannier90_2020}. \\

\medskip
\noindent\textbf{Conflict of Interest}\\
The authors declare no conflict of interest.

\medskip
\noindent\textbf{Data Availability Statement} \\
All data needed to evaluate the conclusions in the paper are presented in the paper and the Supplementary Information. All raw data generated during the current study are available from the corresponding authors upon reasonable request.

\medskip
\noindent\textbf{Supporting Information}\\
Supporting Information is available from the Wiley Online Library or from the author.

\medskip
\noindent\textbf{Acknowledgements}\\
The authors acknowledge the invaluable contribution for the material synthesis from Konstancja Gawryluk, the fruitful discussion with Lun-hui Hu, and the allocation of beam time at S{\textmu}S (VMS spectrometer). 
T.Sha. acknowledges support by the National Natural Science Foundation of China (Grant Nos. 12374105, 12350710785, and 12561160109), and 
the Fundamental Research Funds for the Central Universities. J.Z. acknowledges support by the National Natural Science Foundation of China (Grant No.12474239).
\\

\medskip
\noindent\textbf{Author contributions}\\
T.Sha. conceived and led the project. I.P. and D.J.G. synthesized the sample and performed single-crystal XRD measurements. J.Z. calculated the electronic band structures. K.X., M.M., Y.X., and Q.Z. performed the magnetization and electrical transport measurements. 
K.X., T.Shi., and T.Sha performed the {\textmu}SR measurements and analyzed the data.  K.X., T.Shi., and T.Sha wrote the paper with input from all co-authors.

\medskip
%
\bibliographystyle{MSP}

\bibliography{ScIr2-xSix.bib}

@article{Yin2022,
  title     = {Topological kagome magnets and superconductors},
  author    = {Yin, Jia-Xin and Lian, Biao and Hasan, M. Zahid},
  journal   = {Nature},
  year      = {2022},
  volume    = {612},
  number    = {7941},
  pages     = {647--657},
  doi       = {10.1038/s41586-022-05516-0},
  OPTurl       = {https://doi.org/10.1038/s41586-022-05516-0},
  issn      = {1476-4687}
}

@article{Neupert2022,
  author    = {Neupert, Titus and Denner, M. Michael and Yin, Jia-Xin and Thomale, Ronny and Hasan, M. Zahid},
  title     = {Charge order and superconductivity in kagome materials},
  journal   = {Nat. Phys.},
  year      = {2022},
  volume    = {18},
  number    = {2},
  pages     = {137--143},
  doi       = {10.1038/s41567-021-01404-y},
  OPTurl       = {https://doi.org/10.1038/s41567-021-01404-y},
  issn      = {1745-2481}
}

@article{zhu2024anomalous,
	title = {Anomalous properties in normal and superconducting states of {Sc$_2$Ir$_{4-x}$Si$_x$} due to flat band effect driven by spin-orbit coupling},
	volume = {5},
	issn = {2662-4443},
	OPTurl = {https://www.nature.com/articles/s43246-024-00521-4},
	doi = {10.1038/s43246-024-00521-4},
	number = {1},
	urldate = {2025-10-20},
	journal = {Commun. Mater.},
	author = {Zhu, Zhengyan and Wu, Yuxiang and Fan, Shengtai and Fan, Yiliang and Li, Yiwen and Ye, Yongze and Zhu, Xiyu and Zhang, Haijun and Wen, Hai-Hu},
	month = may,
	year = {2024},
	pages = {85},
}

@book{Yaouanc2011,
author = {A. Yaouanc and P. Dalmas de R\'eotier},
title = {Muon Spin Rotation, Relaxation, and Resonance: Applications to Condensed Matter},
publisher = {Oxford University Press},
year= {2011},
address={Oxford},
OPTseries={International Series of Monographs on Physics}
}

@article{Carrington2003,
  author    = {Carrington, A. and Manzano, F.},
  title     = {Magnetic penetration depth of MgB$_2$},
  journal   = {Physica C},
  year      = {2003},
  volume    = {385},
  number    = {1-2},
  pages     = {205--214},
  doi       = {10.1016/j.physc.2002.11.042}
}

@article{Barford1988,
  author    = {Barford, William and Gunn, J. M. F.},
  title     = {The theory of the measurement of the London penetration depth in uniaxial type II superconductors by muon spin rotation},
  journal   = {Physica C},
  year      = {1988},
  volume    = {156},
  number    = {4},
  pages     = {515--522},
  doi       = {10.1016/0921-4534(88)90472-5}
}

@article{Brandt2003,
  author    = {Brandt, E. H.},
  title     = {Properties of the ideal Ginzburg-Landau vortex lattice},
  journal   = {Phys. Rev. B},
  year      = {2003},
  volume    = {68},
  number    = {5},
  pages     = {054506},
  doi       = {10.1103/PhysRevB.68.054506}
}

@incollection{Kubo1967,
  author    = {Kubo, R. and Toyabe, T.},
  title     = {A stochastic model for low-field resonance and relaxation},
  booktitle = {Magnetic Resonance and Relaxation},
  editor    = {Blinc, R.},
  year      = {1967},
  publisher = {North-Holland},
  pages     = {99--107}
}

@article{Suter2012,
  author =        {A. Suter and B. M. Wojek},
  journal =       {Phys. Procedia},
  pages =         {69--73},
  title =         {Musrfit: {A} Free Platform-Independent Framework for
                   $\mu${SR} Data Analysis},
  volume =        {30},
  year =          {2012},
  doi =           {10.1016/j.phpro.2012.04.042},
  issn =          {1875-3892},
  OPTurl =           {http://www.sciencedirect.com/science/article/pii/S187538921201228X},
}

@book{Amato2024,
  address =       {Cham},
  author =        {A. Amato and E. Morenzoni},
  publisher =     {Springer},
  title =         {Introduction to Muon Spin Spectroscopy: Applications
                   to Solid State and Material Sciences},
  year =          {2024},
  doi =           {10.1007/978-3-031-44959-8},
}

@book{Blundell2021,
  address =       {Oxford},
  editor =        {Blundell, Stephen J. and De Renzi, Roberto and
                   Lancaster, Tom and Pratt, Francis L.},
  month =         nov,
  publisher =     {Oxford University Press},
  title =         {Muon {Spectroscopy}: {An} {Introduction}},
  year =          {2021},
  doi =           {10.1093/oso/9780198858959.001.0001},
}

@article{Shang2020,
  author =        {Shang, T. and Smidman, M. and Wang, A. and Chang, L-J and Baines, C. and Lee, M. K. and Nie, Z. Y. and
                   Pang, G. M. and Xie, W. and Jiang, W. B. and Shi, M. and
                   Medarde, M. and Shiroka, T. and Yuan, H. Q.},
  journal =       {Phys. Rev. Lett.},
  month =         {May},
  pages =         {207001},
  publisher =     {American Physical Society},
  title =         {Simultaneous Nodal Superconductivity and Time-Reversal Symmetry Breaking in the Noncentrosymmetric Superconductor {CaPtAs}},
  volume =        {124},
  year =          {2020},
  doi =           {10.1103/PhysRevLett.124.207001},
  OPTurl =        {https://link.aps.org/doi/10.1103/PhysRevLett.124.207001},
}

@article{Shang2022,
  title = {Evidence of unconventional pairing in the quasi-two-dimensional {CuIr$_{2-x}$Ru$_x$Te$_4$} superconductor},
  author = {Shang, T. and Chen, Y. and Xie, W. and Gawryluk, D. J. and Gupta, R. and Khasanov, R. and Zhu, X. Y. and Zhang, H. and Zhen, Z. X. and Yu, B. C. and Zhou, Z. and Xu, Y. and Zhan, Q. F. and Pomjakushina, E. and Yuan, H. Q. and Shiroka, T.},
  journal = {Phys. Rev. B},
  volume = {106},
  issue = {14},
  pages = {144505},
  numpages = {10},
  year = {2022},
  month = {Oct},
  publisher = {American Physical Society},
  doi = {10.1103/PhysRevB.106.144505},
  OPTurl = {https://link.aps.org/doi/10.1103/PhysRevB.106.144505}
}

@article{Bouquet2001,
	author = {Bouquet, F. and Wang, Y. and Fisher, R. A. and Hinks, D. G. and Jorgensen, J. D. and Junod, A. and Phillips,  N. E.},
	title = {Phenomenological two-gap model for the specific heat of {Mg}{B}$_{2}$},
	journal = {Europhys. Lett.},
	year = {2001},
	month = {Dec},
	publisher = {{IOP} Publishing},
	volume = {56},
	number = {6},
	pages = {856},
	doi = {10.1209/epl/i2001-00598-7},
	OPTurl = {https://iopscience.iop.org/article/10.1209/epl/i2001-00598-7/meta},
}

@article{Biswas2021,
	title = {Chiral singlet superconductivity in the weakly correlated metal {LaPt$_3$P}},
	volume = {12},
	issn = {2041-1723},
	OPTurl = {https://www.nature.com/articles/s41467-021-22807-8},
	doi = {10.1038/s41467-021-22807-8},
	number = {1},
	urldate = {2025-10-24},
	journal = {Nat. Commun.},
	author = {Biswas, P. K. and Ghosh, S. K. and Zhao, J. Z. and Mayoh, D. A. and Zhigadlo, N. D. and Xu, Xiaofeng and Baines, C. and Hillier, A. D. and Balakrishnan, G. and Lees, M. R.},
	month = may,
	year = {2021},
	pages = {2504},
}

@article{Yan2024,
	title = {Structural {Chirality} and {Electronic} {Chirality} in {Quantum} {Materials}},
	volume = {54},
	copyright = {http://creativecommons.org/licenses/by/4.0/},
	issn = {1531-7331, 1545-4118},
	OPTurl = {https://www.annualreviews.org/content/journals/10.1146/annurev-matsci-080222-033548},
	doi = {10.1146/annurev-matsci-080222-033548},
	number = {1},
	urldate = {2025-10-28},
	journal = {Annu. Rev. Mater. Res.},
	author = {Yan, Binghai},
	month = aug,
	year = {2024},
	pages = {97--115},
}

@article{Hasan2021,
  author =        {Hasan, M. Zahid and Chang, Guoqing and
                   Belopolski, Ilya and Bian, Guang and Xu, Su-Yang and
                   Yin, Jia-Xin},
  journal =       {Nat. Rev. Mater.},
  month =         apr,
  OPTnumber =        {9},
  pages =         {784--803},
  title =         {Weyl, {Dirac} and high-fold chiral fermions in
                   topological quantum matter},
  volume =        {6},
  year =          {2021},
  doi =           {10.1038/s41578-021-00301-3},
  issn =          {2058-8437},
}

@article{Narang2021,
  author =        {Narang, Prineha and Garcia, Christina A. C. and
                   Felser, Claudia},
  journal =       {Nat. Mater.},
  month =         mar,
  OPTnumber =      {3},
  pages =         {293--300},
  title =         {The topology of electronic band structures},
  volume =        {20},
  year =          {2021},
  doi =           {10.1038/s41563-020-00820-4},
  issn =          {1476-1122},
}

@article{Bradlyn2016,
  author =        {Bradlyn, Barry and Cano, Jennifer and Wang, Zhijun and
                   Vergniory, M. G. and Felser, C. and Cava, R. J. and
                   Bernevig, B. Andrei},
  journal =       {Science},
  month =         aug,
  OPTnumber =        {6299},
  pages =         {aaf5037},
  title =         {Beyond {Dirac} and {Weyl} fermions: {Unconventional}
                   quasiparticles in conventional crystals},
  volume =        {353},
  year =          {2016},
  doi =           {10.1126/science.aaf5037},
  issn =          {0036-8075, 1095-9203},
}

@article{Chang2018,
  author =        {Chang, Guoqing and Wieder, Benjamin J. and
                   Schindler, Frank and Sanchez, Daniel S. and
                   Belopolski, Ilya and Huang, Shin-Ming and
                   Singh, Bahadur and Wu, Di and Chang, Tay-Rong and
                   Neupert, Titus and Xu, Su-Yang and Lin, Hsin and
                   Hasan, M. Zahid},
  journal =       {Nat. Mater.},
  month =         nov,
  OPTnumber =        {11},
  pages =         {978--985},
  title =         {Topological quantum properties of chiral crystals},
  volume =        {17},
  year =          {2018},
  doi =           {10.1038/s41563-018-0169-3},
  issn =          {1476-1122, 1476-4660},
}

@article{Fernando2017,
  author =        {De Juan, Fernando and Grushin, Adolfo G. and
                   Morimoto, Takahiro and Moore, Joel E},
  journal =       {Nat. Commun.},
  month =         jul,
  OPTnumber =        {1},
  pages =         {15995},
  title =         {Quantized circular photogalvanic effect in {Weyl}
                   semimetals},
  volume =        {8},
  year =          {2017},
  doi =           {10.1038/ncomms15995},
  issn =          {2041-1723},
}

@article{Tan2022,
  author =        {Tan, Wei and Jiang, Xiao and Li, Yang and
                   Wu, Xiaoqiang and Wang, Jianfeng and Huang, Bing},
  journal =       {Adv. Funct. Mater.},
  month =         dec,
  OPTnumber =        {49},
  pages =         {2208023},
  title =         {A Unified Understanding of Diverse Spin Textures of
                   {K}ramers-{W}eyl Fermions in Nonmagnetic Chiral
                   Crystals},
  volume =        {32},
  year =          {2022},
  doi =           {10.1002/adfm.202208023},
  issn =          {1616-301X, 1616-3028},
}

@article{Kresse1996kl,
  author =        {Kresse, G. and Furthm\"uller, J.},
  journal =       {Phys. Rev. B},
  month =         {Oct},
  pages =         {11169},
  publisher =     {American Physical Society},
  title =         {Efficient iterative schemes for ab initio
                   total-energy calculations using a plane-wave basis
                   set},
  volume =        {54},
  year =          {1996},
  doi =           {10.1103/PhysRevB.54.11169},
}

@article{Kresse1996vk,
  author =        {Kresse, G. and Furthm{\"u}ller, J.},
  journal =       {Comput. Mater. Sci.},
  month =         {jul},
  OPTnumber =        {1},
  pages =         {15},
  title =         {Efficiency of ab-initio total energy calculations for
                   metals and semiconductors using a plane-wave basis
                   set},
  volume =        {6},
  year =          {1996},
  doi =           {10.1016/0927-0256(96)00008-0},
}

@article{Kresse1999wc,
  author =        {Kresse, G. and Joubert, D.},
  journal =       {Phys. Rev. B},
  month =         {Jan},
  pages =         {1758},
  publisher =     {American Physical Society},
  title =         {From ultrasoft pseudopotentials to the projector
                   augmented-wave method},
  volume =        {59},
  year =          {1999},
  doi =           {10.1103/PhysRevB.59.1758},
}

@article{Blochl1994zz,
  author =        {Bl\"ochl, P. E.},
  journal =       {Phys. Rev. B},
  month =         {Dec},
  pages =         {17953},
  publisher =     {American Physical Society},
  title =         {Projector augmented-wave method},
  volume =        {50},
  year =          {1994},
  doi =           {10.1103/PhysRevB.50.17953},
}

@article{Perdew1996iq,
  author =        {Perdew, J. P. and Burke, K. and Ernzerhof, M.},
  journal =       {Phys. Rev. Lett.},
  month =         {Oct},
  pages =         {3865},
  publisher =     {American Physical Society},
  title =         {Generalized Gradient Approximation Made Simple},
  volume =        {77},
  year =          {1996},
  doi =           {10.1103/PhysRevLett.77.3865},
}

@article{pizzi_wannier90_2020,
  author =        {Pizzi, Giovanni and Vitale, Valerio and
                   Arita, Ryotaro and Bl\"{u}gel, Stefan and
                   Freimuth, Frank and G\'{e}ranton, Guillaume and
                   Gibertini, Marco and Gresch, Dominik and
                   Johnson, Charles and Koretsune, Takashi and
                   Iba\~{n}ez-Azpiroz, Julen and Lee, Hyungjun and
                   Lihm, Jae-Mo and Marchand, Daniel and
                   Marrazzo, Antimo and Mokrousov, Yuriy and
                   Mustafa, Jamal I and Nohara, Yoshiro and
                   Nomura, Yusuke and Paulatto, Lorenzo and
                   Ponc\'{e}, Samuel and Ponweiser, Thomas and
                   Qiao, Junfeng and Th\"{o}le, Florian and
                   Tsirkin, Stepan S and Wierzbowska, Ma\l{l}gorzata and
                   Marzari, Nicola and Vanderbilt, David and Souza, Ivo and
                   Mostofi, Arash A and Yates, Jonathan R},
  journal =       {J. Phys.: Condens. Matter},
  month =         apr,
  OPTnumber =        {16},
  pages =         {165902},
  title =         {Wannier90 as a community code: new features and
                   applications},
  volume =        {32},
  year =          {2020},
  doi =           {10.1088/1361-648X/ab51ff},
  issn =          {0953-8984, 1361-648X},
}

@article{Prozorov2006,
	author = {Ruslan Prozorov and Russell W. Giannetta},
	title = {Magnetic penetration depth in unconventional superconductors},
	journal = {Supercond. Sci. Technol.},
	volume = {19},
	issue = {8},
	pages = {R41--R67},
	year = {2006},
	month = {Jun},
	publisher = {{IOP} {P}ublishing},
	doi = {10.1088/0953-2048/19/8/r01},
	OPTurl = {https://doi.org/10.1088%2F0953-2048%2F19%2F8%2Fr01},
}

@book{tinkham1996,
  author    = {Tinkham, M.},
  title     = {Introduction to Superconductivity},
  edition   = {2},
  publisher = {Dover Publications},
  address   = {Mineola, NY},
  year      = {1996},
  isbn      = {978-0-486-43503-9}
}

@article{Mielke2021,
  title = {Nodeless kagome superconductivity in {LaRu$_3$Si$_2$}},
  author = {Mielke, C. and Qin, Y. and Yin, J.-X. and Nakamura, H. and Das, D. and Guo, K. and Khasanov, R. and Chang, J. and Wang, Z. Q. and Jia, S. and Nakatsuji, S. and Amato, A. and Luetkens, H. and Xu, G. and Hasan, M. Z. and Guguchia, Z.},
  journal = {Phys. Rev. Mater.},
  volume = {5},
  issue = {3},
  pages = {034803},
  numpages = {7},
  year = {2021},
  month = {Mar},
  publisher = {American Physical Society},
  doi = {10.1103/PhysRevMaterials.5.034803},
  OPTurl = {https://link.aps.org/doi/10.1103/PhysRevMaterials.5.034803},
}

@article{Mielke2022,
	title = {Local spectroscopic evidence for a nodeless magnetic kagome superconductor {CeRu}$_{\textrm{2}}$},
	volume = {34},
	issn = {0953-8984, 1361-648X},
	OPTurl = {https://iopscience.iop.org/article/10.1088/1361-648X/ac9813},
	doi = {10.1088/1361-648X/ac9813},
	number = {48},
	urldate = {2025-11-02},
	journal = {J. Phys.: Condens. Matter},
	author = {Mielke Iii, C and Liu, H and Das, D and Yin, J-X and Deng, L Z and Spring, J and Gupta, R and Medarde, M and Chu, C-W and Khasanov, R and Hasan, Z M and Shi, Y and Luetkens, H and Guguchia, Z},
	month = nov,
	year = {2022},
	pages = {485601},
}

@article{Mielke2022b,
	title = {Time-reversal symmetry-breaking charge order in a kagome superconductor},
	volume = {602},
	copyright = {2022 The Author(s), under exclusive licence to Springer Nature Limited},
	issn = {1476-4687},
	OPTurl = {https://www.nature.com/articles/s41586-021-04327-z},
	doi = {10.1038/s41586-021-04327-z},
	number = {7896},
	urldate = {2025-11-02},
	journal = {Nature},
	author = {Mielke, C. and Das, D. and Yin, J.-X. and Liu, H. and Gupta, R. and Jiang, Y.-X. and Medarde, M. and Wu, X. and Lei, H. C. and Chang, J. and Dai, Pengcheng and Si, Q. and Miao, H. and Thomale, R. and Neupert, T. and Shi, Y. and Khasanov, R. and Hasan, M. Z. and Luetkens, H. and Guguchia, Z.},
	month = feb,
	year = {2022},
	pages = {245--250},
}

@article{Guguchia2023b,
	title = {Tunable unconventional kagome superconductivity in charge ordered {RbV$_3$Sb$_5$} and {KV$_3$Sb$_5$}},
	volume = {14},
	copyright = {2023 The Author(s)},
	issn = {2041-1723},
	OPTurl = {https://www.nature.com/articles/s41467-022-35718-z},
	doi = {10.1038/s41467-022-35718-z},
	number = {1},
	urldate = {2025-11-02},
	journal = {Nat. Commun.},
	author = {Guguchia, Z. and Mielke, C. and Das, D. and Gupta, R. and Yin, J.-X. and Liu, H. and Yin, Q. and Christensen, M. H. and Tu, Z. and Gong, C. and Shumiya, N. and Hossain, Md Shafayat and Gamsakhurdashvili, Ts and Elender, M. and Dai, Pengcheng and Amato, A. and Shi, Y. and Lei, H. C. and Fernandes, R. M. and Hasan, M. Z. and Luetkens, H. and Khasanov, R.},
	month = jan,
	year = {2023},
	pages = {153},
}

@article{Gupta2022,
	title = {Microscopic evidence for anisotropic multigap superconductivity in the {CsV$_3$Sb$_5$} kagome superconductor},
	volume = {7},
	copyright = {2022 The Author(s)},
	issn = {2397-4648},
	OPTurl = {https://www.nature.com/articles/s41535-022-00453-7},
	doi = {10.1038/s41535-022-00453-7},
	number = {1},
	urldate = {2025-11-02},
	journal = {npj Quantum Mater.},
	author = {Gupta, Ritu and Das, Debarchan and Mielke III, Charles Hillis and Guguchia, Zurab and Shiroka, Toni and Baines, Christopher and Bartkowiak, Marek and Luetkens, Hubertus and Khasanov, Rustem and Yin, Qiangwei and Tu, Zhijun and Gong, Chunsheng and Lei, Hechang},
	month = apr,
	year = {2022},
	pages = {49},
}

@article{Gupta2022b,
	title = {Two types of charge order with distinct interplay with superconductivity in the kagome material {CsV$_3$Sb$_5$}},
	volume = {5},
	copyright = {2022 The Author(s)},
	issn = {2399-3650},
	OPTurl = {https://www.nature.com/articles/s42005-022-01011-0},
	doi = {10.1038/s42005-022-01011-0},
	number = {1},
	urldate = {2025-11-02},
	journal = {Commun. Phys.},
	author = {Gupta, Ritu and Das, Debarchan and Mielke, Charles and Ritz, Ethan T. and Hotz, Fabian and Yin, Qiangwei and Tu, Zhijun and Gong, Chunsheng and Lei, Hechang and Birol, Turan and Fernandes, Rafael M. and Guguchia, Zurab and Luetkens, Hubertus and Khasanov, Rustem},
	month = sep,
	year = {2022},
	pages = {232},
}

@article{Graham2024,
	title = {Microscopic probing of the superconducting and normal state properties of {Ta$_2$V$_{3.1}$Si$_{0.9}$} by muon spin rotation},
	volume = {5},
	issn = {2662-4443},
	OPTurl = {https://www.nature.com/articles/s43246-024-00666-2},
	doi = {10.1038/s43246-024-00666-2},
	number = {1},
	urldate = {2025-11-02},
	journal = {Commun. Mater.},
	author = {Graham, J. N. and Liu, H. and Sazgari, V. and Mielke Iii, C. and Medarde, M. and Luetkens, H. and Khasanov, R. and Shi, Y. and Guguchia, Z.},
	month = oct,
	year = {2024},
	pages = {225},
}

@article{Mandal2022,
  title = {Time-reversal symmetry breaking in frustrated superconductor {Re$_2$Hf}},
  author = {Mandal, Manasi and Kataria, Anshu and Patra, Chandan and Singh, D. and Biswas, P. K. and Hillier, A. D. and Das, Tanmoy and Singh, R. P.},
  journal = {Phys. Rev. B},
  volume = {105},
  issue = {9},
  pages = {094513},
  numpages = {8},
  year = {2022},
  month = {Mar},
  publisher = {American Physical Society},
  doi = {10.1103/PhysRevB.105.094513},
  OPTurl = {https://link.aps.org/doi/10.1103/PhysRevB.105.094513}
}

@article{Mandal2025,
  title = {Time-reversal symmetry breaking in a Re-based kagome lattice superconductor},
  author = {Mandal, Manasi and Kataria, A. and Meena, P. K. and Kushwaha, R. K. and Singh, D. and Biswas, P. K. and Stewart, R. and Hillier, A. D. and Singh, R. P.},
  journal = {Phys. Rev. B},
  volume = {111},
  issue = {5},
  pages = {054511},
  numpages = {10},
  year = {2025},
  month = {Feb},
  publisher = {American Physical Society},
  doi = {10.1103/PhysRevB.111.054511},
  OPTurl = {https://link.aps.org/doi/10.1103/PhysRevB.111.054511}
}

@article{Schoop2015,
  title = {Dirac metal to topological metal transition at a structural phase change in {Au$_2$Pb} and prediction of {$\mathbb{Z}_2$} topology for the superconductor},
  author = {Schoop, Leslie M. and Xie, Lilia S. and Chen, Ru and Gibson, Quinn D. and Lapidus, Saul H. and Kimchi, Itamar and Hirschberger, Max and Haldolaarachchige, Neel and Ali, Mazhar N. and Belvin, Carina A. and Liang, Tian and Neaton, Jeffrey B. and Ong, N. P. and Vishwanath, Ashvin and Cava, R. J.},
  journal = {Phys. Rev. B},
  volume = {91},
  issue = {21},
  pages = {214517},
  numpages = {6},
  year = {2015},
  month = {Jun},
  publisher = {American Physical Society},
  doi = {10.1103/PhysRevB.91.214517},
  OPTurl = {https://link.aps.org/doi/10.1103/PhysRevB.91.214517}
}

@article{Xing2016,
	title = {Superconductivity in topologically nontrivial material {Au$_2$Pb}},
	volume = {1},
	copyright = {2016 The Author(s)},
	issn = {2397-4648},
	OPTurl = {https://www.nature.com/articles/npjquantmats20165},
	doi = {10.1038/npjquantmats.2016.5},
	number = {1},
	urldate = {2025-11-03},
	journal = {npj Quantum Mater.},
	author = {Xing, Ying and Wang, He and Li, Chao-Kai and Zhang, Xiao and Liu, Jun and Zhang, Yangwei and Luo, Jiawei and Wang, Ziqiao and Wang, Yong and Ling, Langsheng and Tian, Mingliang and Jia, Shuang and Feng, Ji and Liu, Xiong-Jun and Wei, Jian and Wang, Jian},
	month = jul,
	year = {2016},
	pages = {16005},
}

@article{Geballe1965,
  title = {Superconductivity in Binary Alloy Systems of the Rare Earths and of Thorium with {Pt}-Group Metals},
  author = {Geballe, T. H. and Matthias, B. T. and Compton, V. B. and Corenzwit, E. and Hull, G. W. and Longinotti, L. D.},
  journal = {Phys. Rev.},
  volume = {137},
  issue = {1A},
  pages = {A119--A127},
  numpages = {0},
  year = {1965},
  month = {Jan},
  publisher = {American Physical Society},
  doi = {10.1103/PhysRev.137.A119},
  OPTurl = {https://link.aps.org/doi/10.1103/PhysRev.137.A119}
}

@article{Chowdhury2019,
	title = {An ab-initio {Investigation}: {The} {Physical} {Properties} of {ScIr$_2$} {Superconductor}},
	volume = {61},
	issn = {1090-6460},
	OPTurl = {https://doi.org/10.1134/S1063783419040310},
	doi = {10.1134/S1063783419040310},
	number = {4},
	urldate = {2025-11-03},
	journal = {Phys. Solid State},
	author = {Chowdhury, Uttam Kumar and Saha, Tapas Chandra},
	month = apr,
	year = {2019},
	pages = {530--536},
}

@article{Uzunok2020,
	title = {The {Effect} of {Spin}-{Orbit} {Interaction} {On} {Structural} and {Electronic} {Properties} of {ScIr$_2$}},
	volume = {24},
	issn = {2147-835X},
	OPTurl = {https://dergipark.org.tr/en/pub/saufenbilder/issue/52471/680230},
	doi = {10.16984/saufenbilder.680230},
	number = {2},
	urldate = {2025-11-03},
	journal = {Sakarya Univ. J. Sci.},
	author = {Uzunok, Huseyin Yasin},
	month = apr,
	year = {2020},
	pages = {406--411},
}

@article{Shang2025,
	title = {Discovery of {Nodal}‐{Line} {Superconductivity} in {Chiral} {Crystals}},
	volume = {37},
	issn = {0935-9648, 1521-4095},
	OPTurl = {https://advanced.onlinelibrary.wiley.com/doi/10.1002/adma.202511385},
	doi = {10.1002/adma.202511385},
	number = {43},
	urldate = {2025-11-03},
	journal = {Adv. Mater.},
	author = {Shang, Tian and Zhao, Jianzhou and Hu, Lun‐Hui and Wu, Weikang and Xia, Keqi and Ajeesh, Mukkattu O. and Nicklas, Michael and Xu, Yang and Zhan, Qingfeng and Gawryluk, Dariusz J. and Shi, Ming and Shiroka, Toni},
	month = oct,
	year = {2025},
	pages = {e11385},
}

@article{Carnicom2018,
  author =        {Carnicom, E. M. and Xie, W. and Klimczuk, T. and
                   Lin, J. J. and G{\'o}rnicka, K. and Sobczak, Z. and
                   Ong, N. P. and Cava, R. J.},
  journal =       {Sci. Adv.},
  OPTnumber =        {5},
  pages =         {eaar7969},
  publisher =     {American Association for the Advancement of Science},
  title =         {{Ta}{Rh}$_2${B}$_2$ and {Nb}{Rh}$_2${B}$_2$:
                   {S}uperconductors with a chiral noncentrosymmetric
                   crystal structure},
  volume =        {4},
  year =          {2018},
  doi =           {10.1126/sciadv.aar7969},
}

@article{Sun2015,
  author =        {Sun, Z. X. and Enayat, M. and Maldonado, A. and
                   Lithgow, C. and Yelland, E. and Peets, D. C. and
                   Yaresko, A. and Schnyder, A. P. and Wahl, P.},
  journal =       {Nat. Commun.},
  pages =         {6633},
  publisher =     {Nature Publishing Group},
  title =         {Dirac surface states and nature of superconductivity
                   in noncentrosymmetric {Bi}{Pd}},
  volume =        {6},
  year =          {2015},
  doi =           {10.1038/ncomms7633},
}

@article{yuan2006,
  author =        {Yuan, H. Q. and Agterberg, D. F. and Hayashi, N. and
                   Badica, P. and Vandervelde, D. and Togano, K. and
                   Sigrist, M. and Salamon, M. B.},
  journal =       {Phys. Rev. Lett.},
  month =         {Jul},
  pages =         {017006},
  publisher =     {American Physical Society},
  title =         {{$s$}-Wave Spin-Triplet Order in Superconductors
                   without Inversion Symmetry: {Li}$_{2}${Pd}$_{3}${B}
                   and {Li}$_{2}${Pt}$_{3}${B}},
  volume =        {97},
  year =          {2006},
  doi =           {10.1103/PhysRevLett.97.017006},
}

@article{nishiyama2007,
  author =        {Nishiyama, M. and Inada, Y. and Zheng, G.-q.},
  journal =       {Phys. Rev. Lett.},
  month =         {Jan},
  pages =         {047002},
  publisher =     {American Physical Society},
  title =         {Spin triplet superconducting state due to broken
                   inversion symmetry in {Li}$_{2}${Pt}$_{3}${B}},
  volume =        {98},
  year =          {2007},
  doi =           {10.1103/PhysRevLett.98.047002},
}

@article{Karki2010,
  author =        {Karki, A. B. and Xiong, Y. M. and Vekhter, I. and
                   Browne, D. and Adams, P. W. and Young, D. P. and
                   Thomas, K. R. and Chan, Julia Y. and Kim, H. and
                   Prozorov, R.},
  journal =       {Phys. Rev. B},
  month =         {Aug},
  pages =         {064512},
  publisher =     {American Physical Society},
  title =         {Structure and physical properties of the
                   noncentrosymmetric superconductor {Mo$_3$Al$_2$C}},
  volume =        {82},
  year =          {2010},
  doi =           {10.1103/PhysRevB.82.064512},
}

@article{Amon2018,
  author =        {Amon, A. and Svanidze, E. and Cardoso-Gil, R. and
                   Wilson, M. N. and Rosner, H. and Bobnar, M. and
                   Schnelle, W. and Lynn, J. W. and Gumeniuk, R. and
                   Hennig, C. and Luke, G. M. and Borrmann, H. and
                   Leithe-Jasper, A. and Grin, Yu.},
  journal =       {Phys. Rev. B},
  month =         {Jan},
  pages =         {014501},
  publisher =     {American Physical Society},
  title =         {Noncentrosymmetric superconductor {BeAu}},
  volume =        {97},
  year =          {2018},
  doi =           {10.1103/PhysRevB.97.014501},
}

@article{Tsvyashchenko2016,
  author =        {Tsvyashchenko, A.V. and Sidorov, V.A. and
                   Petrova, A.E. and Fomicheva, L.N. and Zibrov, I.P. and
                   Dmitrienko, V.E.},
  journal =       {J. Alloy. Compd.},
  month =         nov,
  pages =         {431--437},
  title =         {Superconductivity and magnetism in noncentrosymmetric
                   {RhGe}},
  volume =        {686},
  year =          {2016},
  doi =           {10.1016/j.jallcom.2016.06.048},
  issn =          {09258388},
}

@article{Joshi2015,
  author =        {Joshi, B and Thamizhavel, A and Ramakrishnan, S},
  journal =       {J. Phys.: Conf. Ser.},
  month =         mar,
  pages =         {012069},
  title =         {Superconductivity in cubic noncentrosymmetric
                   {PdBiSe} Crystal},
  volume =        {592},
  year =          {2015},
  doi =           {10.1088/1742-6596/592/1/012069},
  issn =          {1742-6596},
}

@article{Smidman2017,
  author =        {Smidman, M. and Salamon, M. B. and Yuan, H. Q. and
                   Agterberg, D. F.},
  journal =       {Rep. Prog. Phys.},
  optnumber =        {3},
  pages =         {036501},
  publisher =     {IOP Publishing},
  title =         {Superconductivity and spin--orbit coupling in
                   non-centrosymmetric materials: {A} review},
  volume =        {80},
  year =          {2017},
}

@article{Gao2022,
  author =        {Gao, Zhe Shen and Gao, Xue-Jian and He, Wen-Yu and
                   Xu, Xiao Yan and Ng, T. K. and Law, K. T.},
  journal =       {Quantum Front.},
  month =         sep,
  OPTnumber =        {1},
  pages =         {3},
  title =         {Topological superconductivity in multifold fermion
                   metals},
  volume =        {1},
  year =          {2022},
  doi =           {10.1007/s44214-022-00001-1},
  issn =          {2731-6106},
}

@article{Koizumi2024,
	title = {Superconductivity in ternary germanite {TaAl}$_x${Ge}$_{2-x}$ with a {C40} chiral structure},
	volume = {63},
	issn = {0021-4922, 1347-4065},
	OPTurl = {https://iopscience.iop.org/article/10.35848/1347-4065/ad920a},
	doi = {10.35848/1347-4065/ad920a},
    number = {12},
	urldate = {2025-11-04},
	journal = {Jpn. J. Appl. Phys.},
	author = {Koizumi, Daigo and Kisanuki, Shingo and Monden, Kenta and Kousaka, Yusuke and Shishido, Hiroaki and Togawa, Yoshihiko},
	month = dec,
	year = {2024},
	pages = {123002},
}

@article{Koshinuma2022,
	title = {High-pressure synthesis and superconductivity of the novel laves phase {BaIr$_2$}},
	volume = {148},
	issn = {09669795},
	OPTurl = {https://linkinghub.elsevier.com/retrieve/pii/S0966979522001844},
	doi = {10.1016/j.intermet.2022.107643},
	urldate = {2025-11-04},
	journal = {Intermetallics},
	author = {Koshinuma, Terunari and Ninomiya, Hiroki and Hase, Izumi and Fujihisa, Hiroshi and Gotoh, Yoshito and Kawashima, Kenji and Ishida, Shigeyuki and Yoshida, Yoshiyuki and Eisaki, Hiroshi and Nishio, Taichiro and Iyo, Akira},
	month = sep,
	year = {2022},
	pages = {107643},
}

@article{Haldo2015,
	title = {Characterization of the heavy metal pyrochlore lattice superconductor {CaIr}$_{\textrm{2}}$},
	volume = {27},
	issn = {0953-8984, 1361-648X},
	OPTurl = {https://iopscience.iop.org/article/10.1088/0953-8984/27/18/185701},
	doi = {10.1088/0953-8984/27/18/185701},
	number = {18},
	urldate = {2025-11-04},
	journal = {J. Phys.: Condens. Matter},
	author = {Haldolaarachchige, Neel and Gibson, Quinn and Schoop, Leslie M and Luo, Huixia and Cava, R J},
	month = may,
	year = {2015},
	pages = {185701},
}

@article{Horie2020,
	title = {Superconductivity in 5 \textit{d} transition metal {Laves} phase {SrIr}$_{\textrm{2}}$},
	volume = {32},
	issn = {0953-8984, 1361-648X},
	OPTurl = {https://iopscience.iop.org/article/10.1088/1361-648X/ab6a2e},
	doi = {10.1088/1361-648X/ab6a2e},
	number = {17},
	urldate = {2025-11-04},
	journal = {J. Phys.: Condens. Matter},
	author = {Horie, R and Horigane, K and Nishiyama, S and Akimitsu, M and Kobayashi, K and Onari, S and Kambe, T and Kubozono, Y and Akimitsu, J},
	month = apr,
	year = {2020},
	pages = {175703},
}

@article{Xiao2021,
	title = {Normal-state and superconducting properties of the cubic {Laves} phase {ThIr$_2$}},
	volume = {128},
	issn = {09669795},
	OPTurl = {https://linkinghub.elsevier.com/retrieve/pii/S0966979520308979},
	doi = {10.1016/j.intermet.2020.106993},
	urldate = {2025-11-04},
	journal = {Intermetallics},
	author = {Xiao, Guorui and Wu, Siqi and Li, Baizhuo and Liu, Bin and Wu, Jifeng and Cui, Yanwei and Zhu, Qinqing and Cao, Guang-han and Ren, Zhi},
	month = jan,
	year = {2021},
	pages = {106993},
}

@article{Yang2023,
	title = {Superconducting properties of the {C15}-type {Laves} phase {ZrIr}$_{\textrm{2}}$ with an {Ir}-based kagome lattice},
	volume = {32},
	issn = {1674-1056},
	OPTurl = {https://iopscience.iop.org/article/10.1088/1674-1056/aca3a2},
	doi = {10.1088/1674-1056/aca3a2},
	number = {1},
	urldate = {2025-11-04},
	journal = {Chin. Phys. B},
	author = {Yang, Qing-Song and Ruan, Bin-Bin and Zhou, Meng-Hu and Gu, Ya-Dong and Ma, Ming-Wei and Chen, Gen-Fu and Ren, Zhi-An},
	month = jan,
	year = {2023},
	pages = {017402},
}

@article{Wang2023,
	title = {Quantum states and intertwining phases in kagome materials},
	volume = {5},
	issn = {2522-5820},
	OPTurl = {https://www.nature.com/articles/s42254-023-00635-7},
	doi = {10.1038/s42254-023-00635-7},
	number = {11},
	urldate = {2025-11-06},
	journal = {Nat. Rev. Phys.},
	author = {Wang, Yaojia and Wu, Heng and McCandless, Gregory T. and Chan, Julia Y. and Ali, Mazhar N.},
	month = sep,
	year = {2023},
	pages = {635--658},
}

@article{He2025,
	title = {Kagom\'{e} lattice compounds and their related systems},
	volume = {61},
	issn = {1359-7345, 1364-548X},
	OPTurl = {https://xlink.rsc.org/?DOI=D4CC06828A},
	doi = {10.1039/D4CC06828A},
	number = {34},
	urldate = {2025-11-06},
	journal = {Chem. Commun.},
	author = {He, Zhangzhen},
	year = {2025},
	pages = {6260--6274},
}

@article{Wilson2024,
	title = {{AV$_3$Sb$_5$} kagome superconductors},
	volume = {9},
	issn = {2058-8437},
	OPTurl = {https://www.nature.com/articles/s41578-024-00677-y},
	doi = {10.1038/s41578-024-00677-y},
	number = {6},
	urldate = {2025-11-06},
	journal = {Nat. Rev. Mater.},
	author = {Wilson, Stephen D. and Ortiz, Brenden R.},
	month = may,
	year = {2024},
	pages = {420--432},
}

@article{Checkelsky2024,
	title = {Flat bands, strange metals and the {Kondo} effect},
	volume = {9},
	issn = {2058-8437},
	OPTurl = {https://www.nature.com/articles/s41578-023-00644-z},
	doi = {10.1038/s41578-023-00644-z},
	number = {7},
	urldate = {2025-11-06},
	journal = {Nat. Rev. Mater.},
	author = {Checkelsky, Joseph G. and Bernevig, B. Andrei and Coleman, Piers and Si, Qimiao and Paschen, Silke},
	month = feb,
	year = {2024},
	pages = {509--526},
}

@article{Guguchia2023,
	title = {Unconventional charge order and superconductivity in kagome-lattice systems as seen by muon-spin rotation},
	volume = {8},
	issn = {2397-4648},
	OPTurl = {https://www.nature.com/articles/s41535-023-00574-7},
	doi = {10.1038/s41535-023-00574-7},
	number = {1},
	urldate = {2025-11-06},
	journal = {npj Quantum Mater.},
	author = {Guguchia, Z. and Khasanov, R. and Luetkens, H.},
	month = aug,
	year = {2023},
	pages = {41},
}

@article{Wang2025,
	title = {Intriguing kagome topological materials},
	volume = {10},
	issn = {2397-4648},
	OPTurl = {https://www.nature.com/articles/s41535-025-00790-3},
	doi = {10.1038/s41535-025-00790-3},
	number = {1},
	urldate = {2025-11-06},
	journal = {npj Quantum Mater.},
	author = {Wang, Qi and Lei, Hechang and Qi, Yanpeng and Felser, Claudia},
	month = jul,
	year = {2025},
	pages = {72},
}

@article{Zhou2017,
  title = {Quantum spin liquid states},
  author = {Zhou, Yi and Kanoda, Kazushi and Ng, Tai-Kai},
  journal = {Rev. Mod. Phys.},
  volume = {89},
  issue = {2},
  pages = {025003},
  numpages = {50},
  year = {2017},
  month = {Apr},
  publisher = {American Physical Society},
  doi = {10.1103/RevModPhys.89.025003},
  OPTurl = {https://link.aps.org/doi/10.1103/RevModPhys.89.025003}
}

@article{Broholm2020,
	title = {Quantum spin liquids},
	volume = {367},
	issn = {0036-8075, 1095-9203},
	OPTurl = {https://www.science.org/doi/10.1126/science.aay0668},
	doi = {10.1126/science.aay0668},
	number = {6475},
	urldate = {2025-11-07},
	journal = {Science},
	author = {Broholm, C. and Cava, R. J. and Kivelson, S. A. and Nocera, D. G. and Norman, M. R. and Senthil, T.},
	month = jan,
	year = {2020},
	pages = {eaay0668},
}

@article{Balents2010,
	title = {Spin liquids in frustrated magnets},
	volume = {464},
	issn = {0028-0836, 1476-4687},
	OPTurl = {https://www.nature.com/articles/nature08917},
	doi = {10.1038/nature08917},
	number = {7286},
	urldate = {2025-11-07},
	journal = {Nature},
	author = {Balents, Leon},
	month = mar,
	year = {2010},
	pages = {199--208},
}

@article{Savary2017,
	title = {Quantum spin liquids: a review},
	volume = {80},
	issn = {0034-4885, 1361-6633},
	shorttitle = {Quantum spin liquids},
	OPTurl = {https://iopscience.iop.org/article/10.1088/0034-4885/80/1/016502},
	doi = {10.1088/0034-4885/80/1/016502},
	number = {1},
	urldate = {2025-11-07},
	journal = {Rep. Prog. Phys.},
	author = {Savary, Lucile and Balents, Leon},
	month = jan,
	year = {2017},
	pages = {016502},
}

@article{Jiang2023,
	title = {Kagome superconductors {AV$_3$Sb$_5$} ({A} = {K}, {Rb}, {Cs})},
	volume = {10},
	copyright = {https://creativecommons.org/licenses/by/4.0/},
	issn = {2095-5138, 2053-714X},
	OPTurl = {https://academic.oup.com/nsr/article/doi/10.1093/nsr/nwac199/6724256},
	doi = {10.1093/nsr/nwac199},
	number = {2},
	urldate = {2025-11-07},
	journal = {Natl. Sci. Rev.},
	author = {Jiang, Kun and Wu, Tao and Yin, Jia-Xin and Wang, Zhenyu and Hasan, M Zahid and Wilson, Stephen D and Chen, Xianhui and Hu, Jiangping},
	month = feb,
	year = {2023},
	pages = {nwac199},
}

@article{Mu2021,
	title = {{S}-{Wave} {Superconductivity} in {Kagome} {Metal} {CsV}$_{\textrm{3}}${Sb}$_{\textrm{5}}$ {Revealed} by $^{\textrm{121/123}}${Sb} {NQR} and $^{\textrm{51}}${V} {NMR} {Measurements}},
	volume = {38},
	issn = {0256-307X, 1741-3540},
	OPTurl = {https://iopscience.iop.org/article/10.1088/0256-307X/38/7/077402},
	doi = {10.1088/0256-307X/38/7/077402},
	number = {7},
	urldate = {2025-11-07},
	journal = {Chin. Phys. Lett.},
	author = {Mu, Chao and Yin, Qiangwei and Tu, Zhijun and Gong, Chunsheng and Lei, Hechang and Li, Zheng and Luo, Jianlin},
	month = jul,
	year = {2021},
	pages = {077402},
}

@article{Nakatsuji2015,
	title = {Large anomalous {Hall} effect in a non-collinear antiferromagnet at room temperature},
	volume = {527},
	issn = {0028-0836, 1476-4687},
	OPTurl = {https://www.nature.com/articles/nature15723},
	doi = {10.1038/nature15723},
	number = {7577},
	urldate = {2025-11-08},
	journal = {Nature},
	author = {Nakatsuji, Satoru and Kiyohara, Naoki and Higo, Tomoya},
	month = nov,
	year = {2015},
	pages = {212--215},
}

@article{Liu2018,
	title = {Giant anomalous {Hall} effect in a ferromagnetic kagome-lattice semimetal},
	volume = {14},
	issn = {1745-2473, 1745-2481},
	OPTurl = {https://www.nature.com/articles/s41567-018-0234-5},
	doi = {10.1038/s41567-018-0234-5},
	number = {11},
	urldate = {2025-11-08},
	journal = {Nat. Phys.},
	author = {Liu, Enke and Sun, Yan and Kumar, Nitesh and Muechler, Lukas and Sun, Aili and Jiao, Lin and Yang, Shuo-Ying and Liu, Defa and Liang, Aiji and Xu, Qiunan and Kroder, Johannes and S\"{u}\ss, Vicky and Borrmann, Horst and Shekhar, Chandra and Wang, Zhaosheng and Xi, Chuanying and Wang, Wenhong and Schnelle, Walter and Wirth, Steffen and Chen, Yulin and Goennenwein, Sebastian T. B. and Felser, Claudia},
	month = nov,
	year = {2018},
	pages = {1125--1131},
}

@article{Negi2025,
	title = {Magnetic {Kagome} materials: bridging fundamental properties and topological quantum applications},
	volume = {12},
	issn = {2051-6347, 2051-6355},
	shorttitle = {Magnetic {Kagome} materials},
	OPTurl = {https://xlink.rsc.org/?DOI=D5MH00120J},
	doi = {10.1039/D5MH00120J},
	number = {13},
	urldate = {2025-11-11},
	journal = {Mater. Horiz.},
	author = {Negi, Pranav and Medhi, Koushik and Pancholi, Abhinav and Roychowdhury, Subhajit},
	year = {2025},
	pages = {4510--4544},
}

@article{Ortiz2019,
title = {New kagome prototype materials: discovery of {KV$_3$Sb$_5$}, {RbV$_3$Sb$_5$}, and {CsV$_3$Sb$_5$}},
  author = {Ortiz, Brenden R. and Gomes, L\'{\i}dia C. and Morey, Jennifer R. and Winiarski, Michal and Bordelon, Mitchell and Mangum, John S. and Oswald, Iain W. H. and Rodriguez-Rivera, Jose A. and Neilson, James R. and Wilson, Stephen D. and Ertekin, Elif and McQueen, Tyrel M. and Toberer, Eric S.},
  journal = {Phys. Rev. Mater.},
  volume = {3},
  issue = {9},
  pages = {094407},
  numpages = {9},
  year = {2019},
  month = {Sep},
  publisher = {American Physical Society},
  doi = {10.1103/PhysRevMaterials.3.094407},
  OPTurl = {https://link.aps.org/doi/10.1103/PhysRevMaterials.3.094407}
}

@article{Ortiz2020,
  title   = {{CsV$_3$Sb$_5$}: A {$\mathbb{Z}_2$} Topological {Kagome} Metal with a Superconducting Ground State},
  author  = {Ortiz, Brenden R. and Teicher, Samuel M. L. and Hu, Yong and Zuo, Julia L. and Sarte, Paul M. and Schueller, Emily C. and Abeykoon, A. M. Milinda and Krogstad, Matthew J. and Rosenkranz, Stephan and Osborn, Raymond and Seshadri, Ram and Balents, Leon and He, Junfeng and Wilson, Stephen D.},
  journal = {Phys. Rev. Lett.},
  year    = {2020},
  volume  = {125},
  number  = {24},
  pages   = {247002},
  doi     = {10.1103/PhysRevLett.125.247002},
  OPurl     = {https://link.aps.org/doi/10.1103/PhysRevLett.125.247002}
}

@article{Jiang2021,
	title = {Unconventional chiral charge order in kagome superconductor {KV$_3$Sb$_5$}},
	volume = {20},
	issn = {1476-1122, 1476-4660},
	OPTurl = {https://www.nature.com/articles/s41563-021-01034-y},
	doi = {10.1038/s41563-021-01034-y},
	number = {10},
	urldate = {2025-11-11},
	journal = {Nat. Mater.},
	author = {Jiang, Yu-Xiao and Yin, Jia-Xin and Denner, M. Michael and Shumiya, Nana and Ortiz, Brenden R. and Xu, Gang and Guguchia, Zurab and He, Junyi and Hossain, Md Shafayat and Liu, Xiaoxiong and Ruff, Jacob and Kautzsch, Linus and Zhang, Songtian S. and Chang, Guoqing and Belopolski, Ilya and Zhang, Qi and Cochran, Tyler A. and Multer, Daniel and Litskevich, Maksim and Cheng, Zi-Jia and Yang, Xian P. and Wang, Ziqiang and Thomale, Ronny and Neupert, Titus and Wilson, Stephen D. and Hasan, M. Zahid},
	month = oct,
	year = {2021},
	pages = {1353--1357},
}

@article{Chen2021,
	title = {Roton pair density wave in a strong-coupling kagome superconductor},
	volume = {599},
	issn = {0028-0836, 1476-4687},
	OPTurl = {https://www.nature.com/articles/s41586-021-03983-5},
	doi = {10.1038/s41586-021-03983-5},
	number = {7884},
	urldate = {2025-11-11},
	journal = {Nature},
	author = {Chen, Hui and Yang, Haitao and Hu, Bin and Zhao, Zhen and Yuan, Jie and Xing, Yuqing and Qian, Guojian and Huang, Zihao and Li, Geng and Ye, Yuhan and Ma, Sheng and Ni, Shunli and Zhang, Hua and Yin, Qiangwei and Gong, Chunsheng and Tu, Zhijun and Lei, Hechang and Tan, Hengxin and Zhou, Sen and Shen, Chengmin and Dong, Xiaoli and Yan, Binghai and Wang, Ziqiang and Gao, Hong-Jun},
	month = nov,
	year = {2021},
	pages = {222--228},
}

@article{Nie2022,
	title = {Charge-density-wave-driven electronic nematicity in a kagome superconductor},
	volume = {604},
	issn = {0028-0836, 1476-4687},
	OPTurl = {https://www.nature.com/articles/s41586-022-04493-8},
	doi = {10.1038/s41586-022-04493-8},
	number = {7904},
	urldate = {2025-11-11},
	journal = {Nature},
	author = {Nie, Linpeng and Sun, Kuanglv and Ma, Wanru and Song, Dianwu and Zheng, Lixuan and Liang, Zuowei and Wu, Ping and Yu, Fanghang and Li, Jian and Shan, Min and Zhao, Dan and Li, Shunjiao and Kang, Baolei and Wu, Zhimian and Zhou, Yanbing and Liu, Kai and Xiang, Ziji and Ying, Jianjun and Wang, Zhenyu and Wu, Tao and Chen, Xianhui},
	month = apr,
	year = {2022},
	pages = {59--64},
}

@article{Liu2023,
	title = {Tunable {Van} {Hove} {Singularity} without {Structural} {Instability} in {Kagome} {Metal} {CsTi}$_3${Bi}$_5$},
	volume = {131},
	issn = {0031-9007, 1079-7114},
	OPTurl = {https://link.aps.org/doi/10.1103/PhysRevLett.131.026701},
	doi = {10.1103/PhysRevLett.131.026701},
	number = {2},
	urldate = {2025-11-13},
	journal = {Phys. Rev. Lett.},
	author = {Liu, Bo and Kuang, Min-Quan and Luo, Yang and Li, Yongkai and Hu, Cheng and Liu, Jiarui and Xiao, Qian and Zheng, Xiquan and Huai, Linwei and Peng, Shuting and Wei, Zhiyuan and Shen, Jianchang and Wang, Bingqian and Miao, Yu and Sun, Xiupeng and Ou, Zhipeng and Cui, Shengtao and Sun, Zhe and Hashimoto, Makoto and Lu, Donghui and Jozwiak, Chris and Bostwick, Aaron and Rotenberg, Eli and Moreschini, Luca and Lanzara, Alessandra and Wang, Yao and Peng, Yingying and Yao, Yugui and Wang, Zhiwei and He, Junfeng},
	month = jul,
	year = {2023},
	pages = {026701},
}

@article{Yang2024,
	title = {Superconductivity and nematic order in a new titanium-based kagome metal {CsTi$_3$Bi$_5$} without charge density wave order},
	volume = {15},
	issn = {2041-1723},
	OPTurl = {https://www.nature.com/articles/s41467-024-53870-6},
	doi = {10.1038/s41467-024-53870-6},
	number = {1},
	urldate = {2025-11-13},
	journal = {Nat. Commun.},
	author = {Yang, Haitao and Ye, Yuhan and Zhao, Zhen and Liu, Jiali and Yi, Xin-Wei and Zhang, Yuhang and Xiao, Hongqin and Shi, Jinan and You, Jing-Yang and Huang, Zihao and Wang, Bingjie and Wang, Jing and Guo, Hui and Lin, Xiao and Shen, Chengmin and Zhou, Wu and Chen, Hui and Dong, Xiaoli and Su, Gang and Wang, Ziqiang and Gao, Hong-Jun},
	month = nov,
	year = {2024},
	pages = {9626},
}

@article{Yang2023b,
	title = {Observation of flat band, {Dirac} nodal lines and topological surface states in {Kagome} superconductor {CsTi$_3$Bi$_5$}},
	volume = {14},
	issn = {2041-1723},
	OPTurl = {https://www.nature.com/articles/s41467-023-39620-0},
	doi = {10.1038/s41467-023-39620-0},
	number = {1},
	urldate = {2025-11-13},
	journal = {Nat. Commun.},
	author = {Yang, Jiangang and Yi, Xinwei and Zhao, Zhen and Xie, Yuyang and Miao, Taimin and Luo, Hailan and Chen, Hao and Liang, Bo and Zhu, Wenpei and Ye, Yuhan and You, Jing-Yang and Gu, Bo and Zhang, Shenjin and Zhang, Fengfeng and Yang, Feng and Wang, Zhimin and Peng, Qinjun and Mao, Hanqing and Liu, Guodong and Xu, Zuyan and Chen, Hui and Yang, Haitao and Su, Gang and Gao, Hongjun and Zhao, Lin and Zhou, X. J.},
	month = jul,
	year = {2023},
	pages = {4089},
}

@article{Li2023,
	title = {Electronic nematicity without charge density waves in titanium-based kagome metal},
	volume = {19},
	issn = {1745-2473, 1745-2481},
	OPTurl = {https://www.nature.com/articles/s41567-023-02176-3},
	doi = {10.1038/s41567-023-02176-3},
	number = {11},
	urldate = {2025-11-13},
	journal = {Nat. Phys.},
	author = {Li, Hong and Cheng, Siyu and Ortiz, Brenden R. and Tan, Hengxin and Werhahn, Dominik and Zeng, Keyu and Johrendt, Dirk and Yan, Binghai and Wang, Ziqiang and Wilson, Stephen D. and Zeljkovic, Ilija},
	month = nov,
	year = {2023},
	pages = {1591--1598},
}

@article{Rehfuss2024,
	title = {Quantum oscillations in kagome metals {CsTi}$_3${Bi}$_5$ and {RbTi}$_3${Bi}$_5$},
	volume = {8},
	issn = {2475-9953},
	OPTurl = {https://link.aps.org/doi/10.1103/PhysRevMaterials.8.024003},
	doi = {10.1103/PhysRevMaterials.8.024003},
	number = {2},
	urldate = {2025-11-13},
	journal = {Phys. Rev. Mater.},
	author = {Rehfuss, Zackary and Broyles, Christopher and Graf, David and Li, Yongkang and Tan, Hengxin and Zhao, Zhen and Liu, Jiali and Zhang, Yuhang and Dong, Xiaoli and Yang, Haitao and Gao, Hongjun and Yan, Binghai and Ran, Sheng},
	month = feb,
	year = {2024},
	pages = {024003},
}

@article{Yi2023,
	title = {Superconducting, {Topological}, and {Transport} {Properties} of {Kagome} {Metals} {CsTi}$_{\textrm{3}}${Bi}$_{\textrm{5}}$ and {RbTi}$_{\textrm{3}}${Bi}$_{\textrm{5}}$},
	volume = {6},
	issn = {2639-5274},
	OPTurl = {https://spj.science.org/doi/10.34133/research.0238},
	doi = {10.34133/research.0238},
	urldate = {2025-11-13},
	journal = {Research},
	author = {Yi, Xin-Wei and Liao, Zheng-Wei and You, Jing-Yang and Gu, Bo and Su, Gang},
	month = jan,
	year = {2023},
	pages = {0238},
}

@article{Wang2025b,
	title = {Spin excitations and flat electronic bands in a {Cr}-based kagome superconductor},
	volume = {16},
	issn = {2041-1723},
	OPTurl = {https://www.nature.com/articles/s41467-025-62298-5},
	doi = {10.1038/s41467-025-62298-5},
	number = {1},
	urldate = {2025-11-13},
	journal = {Nat. Commun.},
	author = {Wang, Zehao and Guo, Yucheng and Huang, Hsiao-Yu and Xie, Fang and Huang, Yuefei and Gao, Bin and Oh, Ji Seop and Wu, Han and Okamoto, Jun and Channagowdra, Ganesha and Chen, Chien-Te and Ye, Feng and Lu, Xingye and Liu, Zhaoyu and Ren, Zheng and Fang, Yuan and Wang, Yiming and Biswas, Ananya and Zhang, Yichen and Yue, Ziqin and Hu, Cheng and Jozwiak, Chris and Bostwick, Aaron and Rotenberg, Eli and Hashimoto, Makoto and Lu, Donghui and Kono, Junichiro and Chu, Jiun-Haw and Yakobson, Boris I. and Birgeneau, Robert J. and Cao, Guang-Han and Fujimori, Atsushi and Huang, Di-Jing and Si, Qimiao and Yi, Ming and Dai, Pengcheng},
	month = aug,
	year = {2025},
	pages = {7573},
}

@article{Liu2024b,
	title = {Superconductivity under pressure in a chromium-based kagome metal},
	volume = {632},
	issn = {0028-0836, 1476-4687},
	OPTurl = {https://www.nature.com/articles/s41586-024-07761-x},
	doi = {10.1038/s41586-024-07761-x},
	number = {8027},
	urldate = {2025-11-13},
	journal = {Nature},
	author = {Liu, Yi and Liu, Zi-Yi and Bao, Jin-Ke and Yang, Peng-Tao and Ji, Liang-Wen and Wu, Si-Qi and Shen, Qin-Xin and Luo, Jun and Yang, Jie and Liu, Ji-Yong and Xu, Chen-Chao and Yang, Wu-Zhang and Chai, Wan-Li and Lu, Jia-Yi and Liu, Chang-Chao and Wang, Bo-Sen and Jiang, Hao and Tao, Qian and Ren, Zhi and Xu, Xiao-Feng and Cao, Chao and Xu, Zhu-An and Zhou, Rui and Cheng, Jin-Guang and Cao, Guang-Han},
	month = aug,
	year = {2024},
	pages = {1032--1037},
}

@article{Li2025,
	title = {Electron correlation and incipient flat bands in the {Kagome} superconductor {CsCr$_3$Sb$_5$}},
	volume = {16},
	issn = {2041-1723},
	OPTurl = {https://www.nature.com/articles/s41467-025-58487-x},
	doi = {10.1038/s41467-025-58487-x},
	number = {1},
	urldate = {2025-11-13},
	journal = {Nat. Commun.},
	author = {Li, Yidian and Liu, Yi and Du, Xian and Wu, Siqi and Zhao, Wenxuan and Zhai, Kaiyi and Hu, Yinqi and Zhang, Senyao and Chen, Houke and Liu, Jieyi and Yang, Yiheng and Peng, Cheng and Hashimoto, Makoto and Lu, Donghui and Liu, Zhongkai and Wang, Yilin and Chen, Yulin and Cao, Guanghan and Yang, Lexian},
	month = apr,
	year = {2025},
	pages = {3229},
}

@article{Wu2025,
	title = {Flat-band enhanced antiferromagnetic fluctuations and superconductivity in pressurized {CsCr$_3$Sb$_5$}},
	volume = {16},
	issn = {2041-1723},
	OPTurl = {https://www.nature.com/articles/s41467-025-56582-7},
	doi = {10.1038/s41467-025-56582-7},
	number = {1},
	urldate = {2025-11-13},
	journal = {Nat. Commun.},
	author = {Wu, Siqi and Xu, Chenchao and Wang, Xiaoqun and Lin, Hai-Qing and Cao, Chao and Cao, Guang-Han},
	month = feb,
	year = {2025},
	pages = {1375},
}

@misc{Yang2022,
	OPTtitle = {Titanium-based kagome superconductor {CsTi}$_3${Bi}$_5$ and topological states},
	OPTurl = {http://arxiv.org/abs/2209.03840},
	doi = {10.48550/arXiv.2209.03840},
	urldate = {2025-11-18},
	publisher = {arXiv},
	author = {Yang, Haitao and Zhao, Zhen and Yi, Xin-Wei and Liu, Jiali and You, Jing-Yang and Zhang, Yuhang and Guo, Hui and Lin, Xiao and Shen, Chengmin and Chen, Hui and Dong, Xiaoli and Su, Gang and Gao, Hong-Jun},
	month = sep,
	year = {2022},
	note = {arXiv:2209.03840},
}

@article{Yin2021,
title = {Strain-sensitive superconductivity in the {kagome} metals {KV$_3$Sb$_5$} and {CsV$_3$Sb$_5$} probed by point-contact spectroscopy},
  author = {Yin, Lichang and Zhang, Dongting and Chen, Chufan and Ye, Ge and Yu, Fanghang and Ortiz, Brenden R. and Luo, Shuaishuai and Duan, Weiyin and Su, Hang and Ying, Jianjun and Wilson, Stephen D. and Chen, Xianhui and Yuan, Huiqiu and Song, Yu and Lu, Xin},
  journal = {Phys. Rev. B},
  volume = {104},
  issue = {17},
  pages = {174507},
  numpages = {7},
  year = {2021},
  month = {Nov},
  publisher = {American Physical Society},
  doi = {10.1103/PhysRevB.104.174507},
  OPTurl = {https://link.aps.org/doi/10.1103/PhysRevB.104.174507}
}

@article{Roppongi2023,
	title = {Bulk evidence of anisotropic s-wave pairing with no sign change in the kagome superconductor {CsV$_3$Sb$_5$}},
	volume = {14},
	issn = {2041-1723},
	OPTurl = {https://www.nature.com/articles/s41467-023-36273-x},
	doi = {10.1038/s41467-023-36273-x},
	number = {1},
	urldate = {2025-11-19},
	journal = {Nat. Commun.},
	author = {Roppongi, M. and Ishihara, K. and Tanaka, Y. and Ogawa, K. and Okada, K. and Liu, S. and Mukasa, K. and Mizukami, Y. and Uwatoko, Y. and Grasset, R. and Konczykowski, M. and Ortiz, B. R. and Wilson, S. D. and Hashimoto, K. and Shibauchi, T.},
	month = feb,
	year = {2023},
	pages = {667},
}

@article{Zhong2023,
	title = {Nodeless electron pairing in {CsV$_3$Sb$_5$}-derived kagome superconductors},
	volume = {617},
	issn = {0028-0836, 1476-4687},
	OPTurl = {https://www.nature.com/articles/s41586-023-05907-x},
	doi = {10.1038/s41586-023-05907-x},
	number = {7961},
	urldate = {2025-11-19},
	journal = {Nature},
	author = {Zhong, Yigui and Liu, Jinjin and Wu, Xianxin and Guguchia, Zurab and Yin, J.-X. and Mine, Akifumi and Li, Yongkai and Najafzadeh, Sahand and Das, Debarchan and Mielke, Charles and Khasanov, Rustem and Luetkens, Hubertus and Suzuki, Takeshi and Liu, Kecheng and Han, Xinloong and Kondo, Takeshi and Hu, Jiangping and Shin, Shik and Wang, Zhiwei and Shi, Xun and Yao, Yugui and Okazaki, Kozo},
	month = may,
	year = {2023},
	pages = {488--492},
}

@article{Duan2021,
	title = {Nodeless superconductivity in the kagome metal {CsV$_3$Sb$_5$}},
	volume = {64},
	issn = {1674-7348, 1869-1927},
	OPTurl = {https://link.springer.com/10.1007/s11433-021-1747-7},
	doi = {10.1007/s11433-021-1747-7},
	number = {10},
	urldate = {2025-11-19},
	journal = {Sci. China Phys. Mech. Astron.},
	author = {Duan, Weiyin and Nie, Zhiyong and Luo, Shuaishuai and Yu, Fanghang and Ortiz, Brenden R. and Yin, Lichang and Su, Hang and Du, Feng and Wang, An and Chen, Ye and Lu, Xin and Ying, Jianjun and Wilson, Stephen D. and Chen, Xianhui and Song, Yu and Yuan, Huiqiu},
	month = oct,
	year = {2021},
	pages = {107462},
}

@article{Shan2022,
  title = {Muon spin relaxation study of the layered kagome superconductor {CsV$_3$Sb$_5$}},
  author = {Shan, Zhaoyang and Biswas, Pabitra K. and Ghosh, Sudeep K. and Tula, T. and Hillier, Adrian D. and Adroja, Devashibhai and Cottrell, Stephen and Cao, Guang-Han and Liu, Yi and Xu, Xiaofeng and Song, Yu and Yuan, Huiqiu and Smidman, Michael},
  journal = {Phys. Rev. Res.},
  volume = {4},
  issue = {3},
  pages = {033145},
  numpages = {8},
  year = {2022},
  month = {Aug},
  publisher = {American Physical Society},
  doi = {10.1103/PhysRevResearch.4.033145},
  OPTurl = {https://link.aps.org/doi/10.1103/PhysRevResearch.4.033145}
}

@article{Xu2021,
  title = {Multiband Superconductivity with Sign-Preserving Order Parameter in {Kagome} Superconductor {CsV$_3$Sb$_5$}},
  author = {Xu, Han-Shu and Yan, Ya-Jun and Yin, Ruotong and Xia, Wei and Fang, Shijie and Chen, Ziyuan and Li, Yuanji and Yang, Wenqi and Guo, Yanfeng and Feng, Dong-Lai},
  journal = {Phys. Rev. Lett.},
  volume = {127},
  issue = {18},
  pages = {187004},
  numpages = {7},
  year = {2021},
  month = {Oct},
  publisher = {American Physical Society},
  doi = {10.1103/PhysRevLett.127.187004},
  OPTurl = {https://link.aps.org/doi/10.1103/PhysRevLett.127.187004}
}

@article{yu_encyclopedia_2022,
	title = {Encyclopedia of emergent particles in three-dimensional crystals},
	volume = {67},
	copyright = {https://www.elsevier.com/tdm/userlicense/1.0/},
	issn = {20959273},
	OPTurl = {https://linkinghub.elsevier.com/retrieve/pii/S2095927321006927},
	doi = {10.1016/j.scib.2021.10.023},
	number = {4},
	urldate = {2024-06-10},
	journal = {Sci. Bull.},
	author = {Yu, Zhi-Ming and Zhang, Zeying and Liu, Gui-Bin and Wu, Weikang and Li, Xiao-Ping and Zhang, Run-Wu and Yang, Shengyuan A. and Yao, Yugui},
	month = feb,
	year = {2022},
	pages = {375}
}

@article{sanchez_topological_2019,
	title = {Topological chiral crystals with helicoid-arc quantum states},
	volume = {567},
	issn = {0028-0836, 1476-4687},
	OPTurl = {https://www.nature.com/articles/s41586-019-1037-2},
	doi = {10.1038/s41586-019-1037-2},
	number = {7749},
	urldate = {2025-12-25},
	journal = {Nature},
	author = {Sanchez, Daniel S. and Belopolski, Ilya and Cochran, Tyler A. and Xu, Xitong and Yin, Jia-Xin and Chang, Guoqing and Xie, Weiwei and Manna, Kaustuv and Süß, Vicky and Huang, Cheng-Yi and Alidoust, Nasser and Multer, Daniel and Zhang, Songtian S. and Shumiya, Nana and Wang, Xirui and Wang, Guang-Qiang and Chang, Tay-Rong and Felser, Claudia and Xu, Su-Yang and Jia, Shuang and Lin, Hsin and Hasan, M. Zahid},
	month = mar,
	year = {2019},
	pages = {500--505}
}

@article{Petricek2020,
	title = {Jana2020 – a new version of the crystallographic computing system {Jana}},
	volume = {238},
	issn = {2194-4946, 2196-7105},
	OPTurl = {https://www.degruyter.com/document/doi/10.1515/zkri-2023-0005/html},
	doi = {10.1515/zkri-2023-0005},
	number = {7-8},
	urldate = {2026-01-09},
	journal = {Z. Kristallogr.},
	author = {Pet\v{r}\'{\i}\v{c}ek, V\'{a}clav and Palatinus, Luk\'{a}\v{s} and Pl\'{a}\v{s}il, Jakub and Du\v{s}ek, Michal},
	month = jul,
	year = {2023},
	pages = {271--282},
}

@article{yang2020b,
	title = {Superconducting behavior of a new metal iridate compound, {Sr}{Ir}$_2$, under pressure},
	volume = {32},
	issn = {0953-8984, 1361-648X},
	OPurl = {https://iopscience.iop.org/article/10.1088/1361-648X/ab4605},
	OPdoi = {10.1088/1361-648X/ab4605},
	number = {2},
	urldate = {2026-01-09},
	journal = {J. Phys. Condens. Matter},
	author = {Yang, Xiaofan and Li, Huan and He, Tong and Taguchi, Tomoya and Wang, Yanan and Goto, Hidenori and Eguchi, Ritsuko and Horie, Rie and Horigane, Kazumasa and Kobayashi, Kaya and Akimitsu, Jun and Ishii, Hirofumi and Liao, Yen-Fa and Yamaoka, Hitoshi and Kubozono, Yoshihiro},
	month = jan,
	year = {2020},
	pages = {025704},
}

@article{li2021,
	title = {Pressure {Dependence} of {Superconducting} {Behavior} of 4d and 5d {Transition} {Metal} {Compounds} {CaRh}$_{\textrm{2}}$ and {CaIr}$_{\textrm{2}}$},
	volume = {125},
	copyright = {https://doi.org/10.15223/policy-029},
	issn = {1932-7447, 1932-7455},
	OPurl = {https://pubs.acs.org/doi/10.1021/acs.jpcc.1c06207},
	OPdoi = {10.1021/acs.jpcc.1c06207},
	language = {en},
	number = {37},
	urldate = {2026-01-09},
	journal = {J. Phys. Chem. C},
	author = {Li, Huan and Taguchi, Tomoya and Wang, Yanan and Goto, Hidenori and Eguchi, Ritsuko and Ishii, Hirofumi and Liao, Yen-Fa and Kubozono, Yoshihiro},
	month = sep,
	year = {2021},
	pages = {20617--20625},
}

@article{Rao2019,
  author =        {Rao, Zhicheng and Li, Hang and Zhang, Tiantian and
                   Tian, Shangjie and Li, Chenghe and Fu, Binbin and
                   Tang, Cenyao and Wang, Le and Li, Zhilin and
                   Fan, Wenhui and Li, Jiajun and Huang, Yaobo and
                   Liu, Zhehong and Long, Youwen and Fang, Chen and
                   Weng, Hongming and Shi, Youguo and Lei, Hechang and
                   Sun, Yujie and Qian, Tian and Ding, Hong},
  journal =       {Nature},
  month =         mar,
  OPTnumber =        {7749},
  pages =         {496--499},
  title =         {Observation of unconventional chiral fermions with
                   long {Fermi} arcs in {CoSi}},
  volume =        {567},
  year =          {2019},
  doi =           {10.1038/s41586-019-1031-8},
  issn =          {0028-0836, 1476-4687},
}

@article{Wu2021,
  title = {Nature of Unconventional Pairing in the Kagome Superconductors $A{\mathrm{V}}_{3}{\mathrm{Sb}}_{5}$ ($A=\mathrm{K},\mathrm{Rb},\mathrm{Cs}$)},
  author = {Wu, Xianxin and Schwemmer, Tilman and M\"uller, Tobias and Consiglio, Armando and Sangiovanni, Giorgio and Di Sante, Domenico and Iqbal, Yasir and Hanke, Werner and Schnyder, Andreas P. and Denner, M. Michael and Fischer, Mark H. and Neupert, Titus and Thomale, Ronny},
  journal = {Phys. Rev. Lett.},
  volume = {127},
  issue = {17},
  pages = {177001},
  numpages = {7},
  year = {2021},
  month = {Oct},
  publisher = {American Physical Society},
  doi = {10.1103/PhysRevLett.127.177001},
  url = {https://link.aps.org/doi/10.1103/PhysRevLett.127.177001}
}

@article{Wen2022,
  title = {Superconducting pairing symmetry in the kagome-lattice Hubbard model},
  author = {Wen, Chenyue and Zhu, Xingchuan and Xiao, Zhisong and Hao, Ning and Mondaini, Rubem and Guo, Huaiming and Feng, Shiping},
  journal = {Phys. Rev. B},
  volume = {105},
  issue = {7},
  pages = {075118},
  numpages = {8},
  year = {2022},
  month = {Feb},
  publisher = {American Physical Society},
  doi = {10.1103/PhysRevB.105.075118},
  url = {https://link.aps.org/doi/10.1103/PhysRevB.105.075118}
}

@article{Romer2022,
  title = {Superconductivity from repulsive interactions on the kagome lattice},
  author = {R\o{}mer, Astrid T. and Bhattacharyya, Shinibali and Valent\'{\i}, Roser and Christensen, Morten H. and Andersen, Brian M.},
  journal = {Phys. Rev. B},
  volume = {106},
  issue = {17},
  pages = {174514},
  numpages = {13},
  year = {2022},
  month = {Nov},
  publisher = {American Physical Society},
  doi = {10.1103/PhysRevB.106.174514},
  url = {https://link.aps.org/doi/10.1103/PhysRevB.106.174514}
}

@book{Bauer2012,
  editor = {E. Bauer and M. Sigrist},
  title = {Non-Centrosymmetric Superconductors},
  publisher = {Springer Verlag},
  year= {2012},
  volume={847},
  address={Berlin},
  OPTseries={Lecture Notes in Physics}
}

@Article{Sungkit2014,
  author    = {Y. Sungkit},
  title     = {Noncentrosymmetric superconductors},
  journal   = {Annu. Rev. Condens. Matter Phys.},
  year      = {2014},
  volume    = {5},
  number    = {2},
  pages     = {15},
  doi       = {10.1146/annurev-conmatphys-031113-133912},
  publisher = {Annual Reviews},
}

@Article{Bonalde2005,
  author    = {Bonalde, I. and Br{\"a}mer-Escamilla, W. and Bauer, E.},
  title     = {Evidence for Line Nodes in the Superconducting Energy Gap of Noncentrosymmetric {Ce}{Pt}$_{3}${Si} from Magnetic Penetration Depth Measurements},
  journal   = {Phys. Rev. Lett.},
  year      = {2005},
  volume    = {94},
  pages     = {207002},
  month     = {May},
  doi       = {10.1103/PhysRevLett.94.207002},
  issue     = {20},
  numpages  = {4},
  publisher = {American Physical Society},
  url       = {https://link.aps.org/doi/10.1103/PhysRevLett.94.207002},
}

@article{Lv2020,
  title = {Type-I superconductivity in noncentrosymmetric $\mathrm{Nb}{\mathrm{Ge}}_{2}$},
  author = {Lv, Baijiang and Li, Miaocong and Chen, Jia and Yang, Yusen and Wu, Siqi and Qiao, Lei and Guan, Feihong and Xing, Hui and Tao, Qian and Cao, Guang-Han and Xu, Zhu-An},
  journal = {Phys. Rev. B},
  volume = {102},
  issue = {6},
  pages = {064507},
  numpages = {7},
  year = {2020},
  month = {Aug},
  publisher = {American Physical Society},
  doi = {10.1103/PhysRevB.102.064507},
  url = {https://link.aps.org/doi/10.1103/PhysRevB.102.064507}
}

@article{Yao2025,
  title = {Observation of chiral surface state in superconducting ${\mathrm{NbGe}}_{2}$},
  author = {Yao, Mengyu and Gutierrez-Amigo, Martin and Roychowdhury, Subhajit and Errea, Ion and Fedorov, Alexander and Strocov, Vladimir N. and Vergniory, Maia G. and Felser, Claudia},
  journal = {Phys. Rev. Mater.},
  volume = {9},
  issue = {3},
  pages = {034803},
  numpages = {7},
  year = {2025},
  month = {Mar},
  publisher = {American Physical Society},
  doi = {10.1103/PhysRevMaterials.9.034803},
  url = {https://link.aps.org/doi/10.1103/PhysRevMaterials.9.034803}
}

@article{Mardanya2024,
  title = {Unconventional superconducting pairing in a {B20} multifold {Weyl} fermion semimetal},
  author = {Mardanya, Sougata and Kargarian, Mehdi and Verma, Rahul and Chang, Tay-Rong and Chowdhury, Sugata and Lin, Hsin and Bansil, Arun and Agarwal, Amit and Singh, Bahadur},
  journal = {Phys. Rev. Mater.},
  volume = {8},
  issue = {9},
  pages = {L091801},
  numpages = {8},
  year = {2024},
  month = {Sep},
  publisher = {American Physical Society},
  doi = {10.1103/PhysRevMaterials.8.L091801},
  url = {https://link.aps.org/doi/10.1103/PhysRevMaterials.8.L091801}
}

@article{Lee2021,
  title = {Topological multiband $s$-wave superconductivity in coupled multifold fermions},
  author = {Lee, Changhee and Yoon, Chiho and Kim, Taehyeok and Chung, Suk Bum and Min, Hongki},
  journal = {Phys. Rev. B},
  volume = {104},
  issue = {24},
  pages = {L241115},
  numpages = {6},
  year = {2021},
  month = {Dec},
  publisher = {American Physical Society},
  doi = {10.1103/PhysRevB.104.L241115},
  url = {https://link.aps.org/doi/10.1103/PhysRevB.104.L241115}
}

@article{Fu2008,
  title = {Superconducting Proximity Effect and Majorana Fermions at the Surface of a Topological Insulator},
  author = {Fu, Liang and Kane, C. L.},
  journal = {Phys. Rev. Lett.},
  volume = {100},
  issue = {9},
  pages = {096407},
  numpages = {4},
  year = {2008},
  month = {Mar},
  publisher = {American Physical Society},
  doi = {10.1103/PhysRevLett.100.096407},
  url = {https://link.aps.org/doi/10.1103/PhysRevLett.100.096407}
}

@article{Qi2011,
  title = {Topological insulators and superconductors},
  author = {Qi, Xiao-Liang and Zhang, Shou-Cheng},
  journal = {Rev. Mod. Phys.},
  volume = {83},
  issue = {4},
  pages = {1057--1110},
  numpages = {0},
  year = {2011},
  month = {Oct},
  publisher = {American Physical Society},
  doi = {10.1103/RevModPhys.83.1057},
  url = {https://link.aps.org/doi/10.1103/RevModPhys.83.1057}
}
\clearpage

\begin{figure*}[!thp]
	\centering
	\includegraphics[width=0.98\textwidth,angle=0]{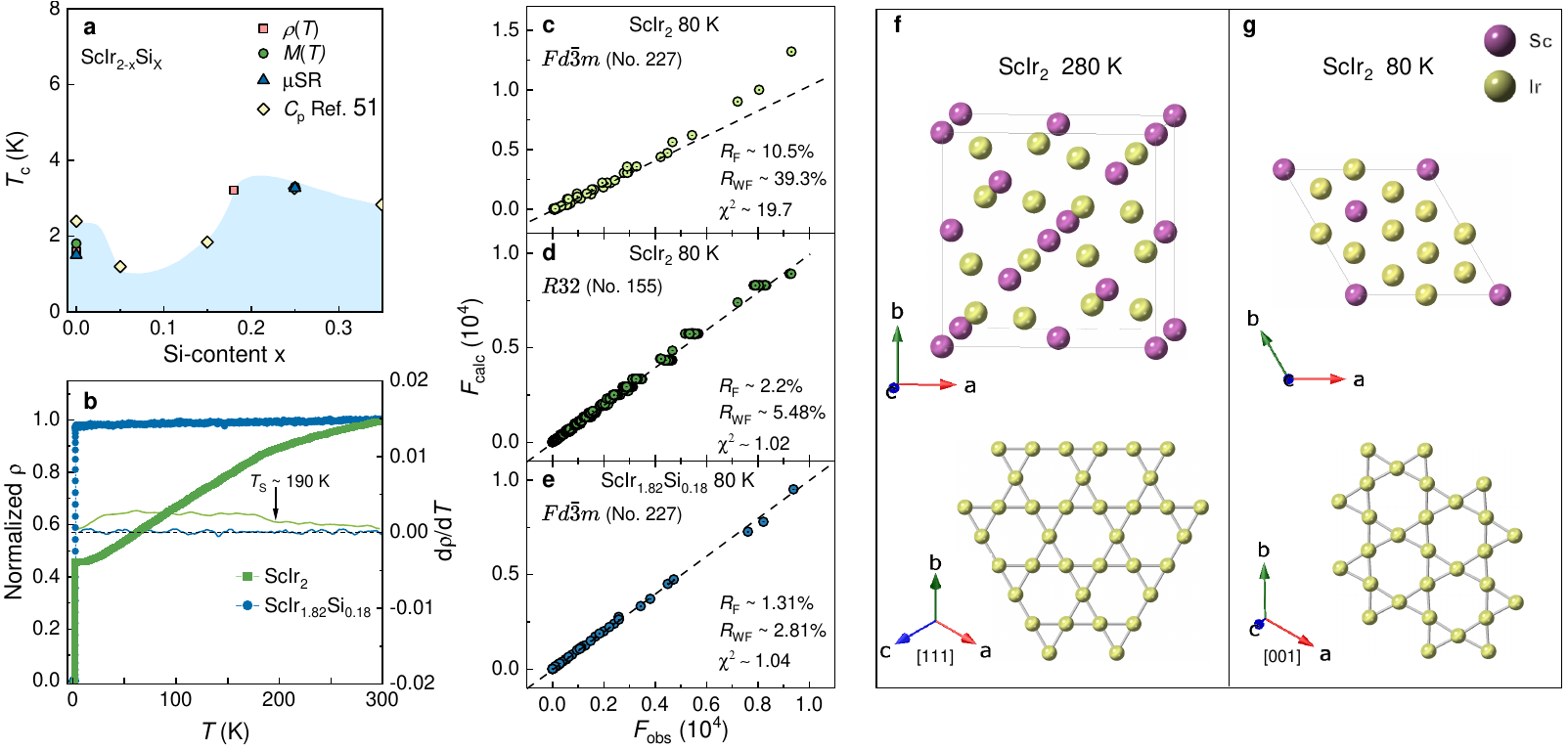}
	\caption{\label{fig:phase} \tcr{
		Phase diagram and crystal structure. 
        a) Superconducting phase diagram of ScIr$_{2-x}$Si$_{x}$ ($0 \le x \le 0.35$). Superconducting transition temperatures determined from the specific heat $C_\mathrm{p}$ were taken from Ref.~\cite{zhu2024anomalous}. Note that, previously the same compound was dubbed Sc$_2$Ir$_{4-x}$Si$_{x}$ ($0 \le x  \le 0.7$)~\cite{zhu2024anomalous}, where the nominal Si concentrations were used. 
		b) Temperature-dependent electrical resistivity $\rho(T)$ (left axis) and its derivative with respect to temperature $\mathrm{d}\rho(T)/\mathrm{d}T$ (right axis) for ScIr$_2$ and ScIr$_{1.82}$Si$_{0.18}$, respectively. For clarity, the electrical resistivity data are normalized to their room-temperature value. The arrow marks a structural phase transition at $T_\mathrm{S}$ $\approx$ 190\,K in ScIr$_2$. 
		c-e) Refinements of SXRD at $T$ = 80\,K of ScIr$_2$ (c,d) and ScIr$_{1.82}$Si$_{0.18}$ (e). For ScIr$_2$, two different structure models, with space groups $Fd\bar{3}m$ (No.~227) (panel c) and $R32$ (No.~155) (panel d), were used for the refinements (see details in Table~S1, Supporting Information).  
		f,g) Crystal structures of the high-$T$ (i.e., $T$ > $T_\mathrm{S}$) cubic phase (f) and the low-$T$ (i.e., $T$ < $T_\mathrm{S}$) rhombohedral phase (g) of ScIr$_2$.   
		In the high-$T$- and the low-$T$ phases, Ir atoms form a kagome layer viewed along the [111]- and the [001]-direction, respectively.}}
	\end{figure*}

\clearpage

\begin{figure}[!thp]
	\centering
	\includegraphics[width=0.48\textwidth,angle= 0]{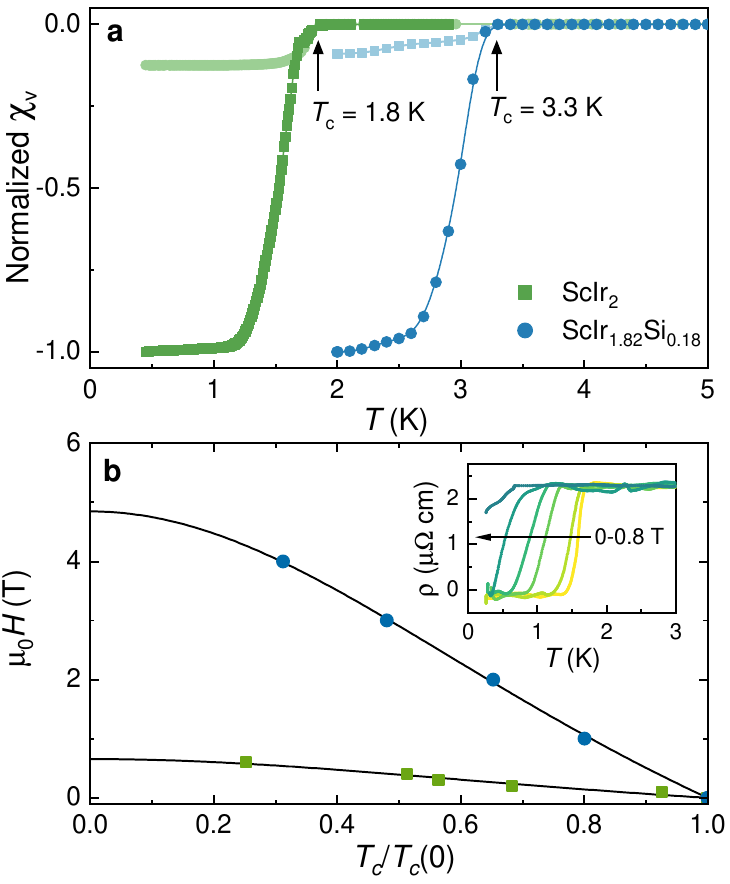}
	\caption{\label{fig:SC} 
		\tcr{
		Characterization of superconductivity. 
		a) Temperature-dependent magnetic susceptibility $\chi(T)$ for ScIr$_2$ and ScIr$_{1.82}$Si$_{0.18}$. Magnetic susceptibility data were collected using both field-cooled (FC) and zero-field-cooled (ZFC) protocols in an applied magnetic field of 1\,mT. Magnetic susceptibility data were normalized to their lowest-temperature value.
		b) Upper critical fields $H_\mathrm{c2}$ of ScIr$_2$ and ScIr$_{1.82}$Si$_{0.18}$. The inset shows the temperature-dependent electrical resistivity $\rho(T)$ collected under different magnetic fields for ScIr$_2$, with ScIr$_{1.82}$Si$_{0.18}$ showing similar features. The symbols in the main panel represent the superconducting transition temperature $T_c$, where the electrical resistivity drops to zero (see inset). 
		The $T_c$s are normalized to the zero-field $T_c(0)$ value. Solid lines are fits to the GL model. }
	}
	\end{figure}
\clearpage

\begin{figure*}[!thp]
	\centering
	\includegraphics[width=1.0\textwidth,angle=0]{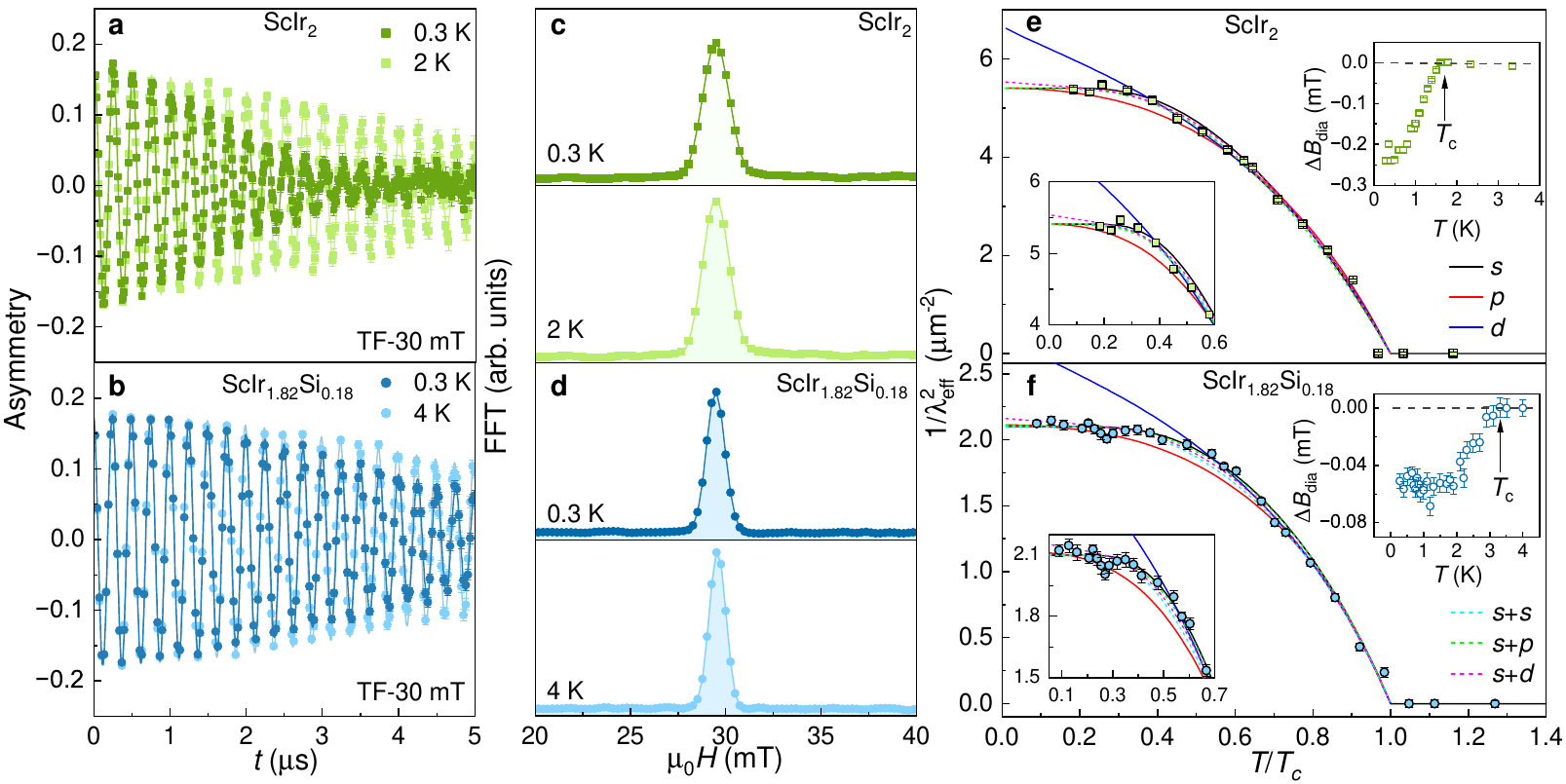}
	\caption{\label{fig:TF-μSR} \tcr{
	Exploring the superconducting pairing via TF-{\textmu}SR.
	a,b) TF-{\textmu}SR spectra collected in the superconducting and normal states in an applied magnetic field of 30\,mT for ScIr$_2$ (a) and ScIr$_{1.82}$Si$_{0.18}$ (b). All the spectra were collected using a FC protocol. Solid lines through the data are fits to Eq.~\eqref{eq:TF}.
	c,d) FFT of the relevant {\textmu}SR spectra in panels (a) and (b) for ScIr$_2$ (c) and ScIr$_{1.82}$Si$_{0.18}$ (d). 
	e,f) Superfluid density [$\rho_\mathrm{sc}(T)$ $\propto$ $\lambda_\mathrm{eff}^{-2}(T)$] as a function of the reduced temperature $T/T_c$ for ScIr$_2$ (e) and ScIr$_{1.82}$Si$_{0.18}$ (f). The upper insets plot temperature-dependent diamagnetic shift $\Delta$$B_\mathrm{dia}$. Here, $\Delta$$B_\mathrm{dia}$ =  $B_\mathrm{s}$ - $B_\mathrm{appl}$, with $B_\mathrm{s}$ the local field sensed by implanted muons in the sample and $B_\mathrm{appl}$ the applied magnetic field. As denoted by the arrows, $\Delta$$B_\mathrm{dia}$ appears below $T_c$ due to the formation of FLL. Lower insets enlarge the low-$T$ data. 
    Lines represent fits to the various models, including single-gap $s$-, $p$-, and $d$-wave (solid black, red, and blue lines), and two-gap $s+s$, $s+p$, $s+d$-wave models (dashed cyan, green, and magenta lines). The derived  fit parameters are summarized in Table~\ref{tab:parameters}.
  	The error bars of $\lambda_\mathrm{eff}^{-2}$ and $\Delta$$B_\mathrm{dia}$ are the SDs obtained from fits of the TF-{\textmu}SR spectra by the \texttt{musrfit} software package~\cite{Suter2012}. }
	}
\end{figure*}
\clearpage

\begin{figure}[!thp]
	\centering
	\includegraphics[width=0.45\textwidth,angle= 0]{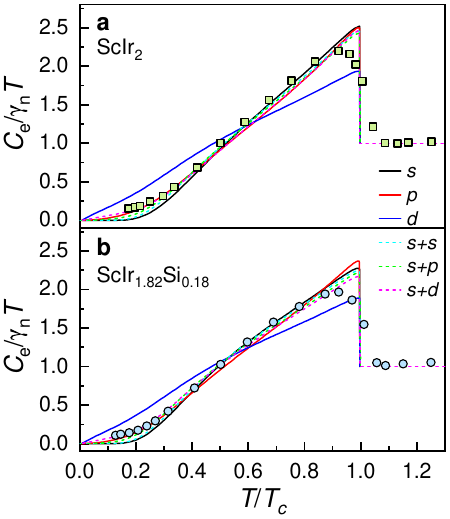}
	\caption{\label{fig_Cp} \tcr{
		Electronic specific heat.
		a,b) Normalized electronic specific heat $C_\mathrm{e}/\gamma_\mathrm{n}T$ as a function of the reduced temperature $T/T_c$ for ScIr$_2$ (a) and ScIr$_{1.82}$Si$_{0.18}$ (b). The different lines represent fits to the various models, analogous to those used to analyze the superfluid density in Figure~\ref{fig:TF-μSR}e,f (see text for details). 
		The derived fit parameters are also listed in Table~\ref{tab:parameters}. Zero-field specific heat data were taken from Ref.~\cite{zhu2024anomalous}. Note that, the extra phases observed in previous studies alsoresulted in lower actual Si concentrations in the doped ScIr$_2$ samples. Since the same method was used to prepare the ScIr$_{1-x}$Si$_{x}$ samples, comparable Si concentrations are expected in both cases. For simplicity, the ScIr$_{1.75}$Si$_{0.25}$ sample is also denoted as ScIr$_{1.82}$Si$_{0.18}$ in the specific-heat results}.
	 }
	\end{figure}
\clearpage

\begin{figure}[!thp]
	\centering
	\includegraphics[width=0.45\textwidth,angle= 0]{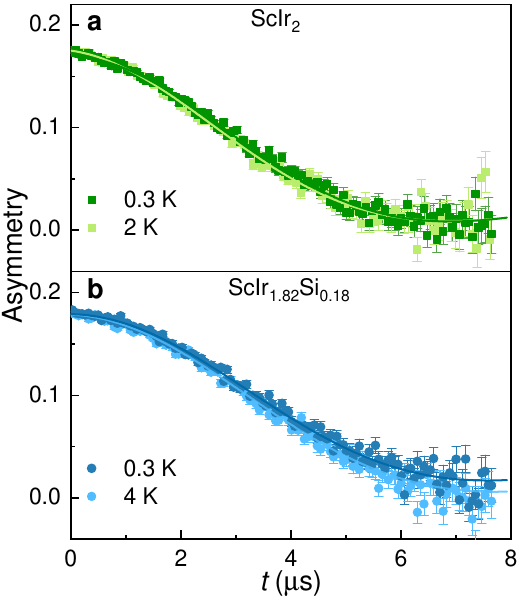}
	\caption{\label{fig:ZF-muSR} \tcr{
		Preserved TRS in the superconducting state of ScIr$_{2-x}$Si$_x$.
		a,b) ZF-{\textmu}SR spectra of ScIr$_2$ (a) and ScIr$_{1.82}$Si$_{0.18}$ (b), collected in their superconducting (0.3\,K) and normal (2 and 4\,K) states. Solid lines through the data are fits to Eq.~\eqref{eq:ZF} using the Kubo-Toyabe relaxation function. 
		}}
\end{figure}
\clearpage

\begin{figure*}[!thp]
	\centering
	\includegraphics[width=0.9\textwidth,angle= 0]{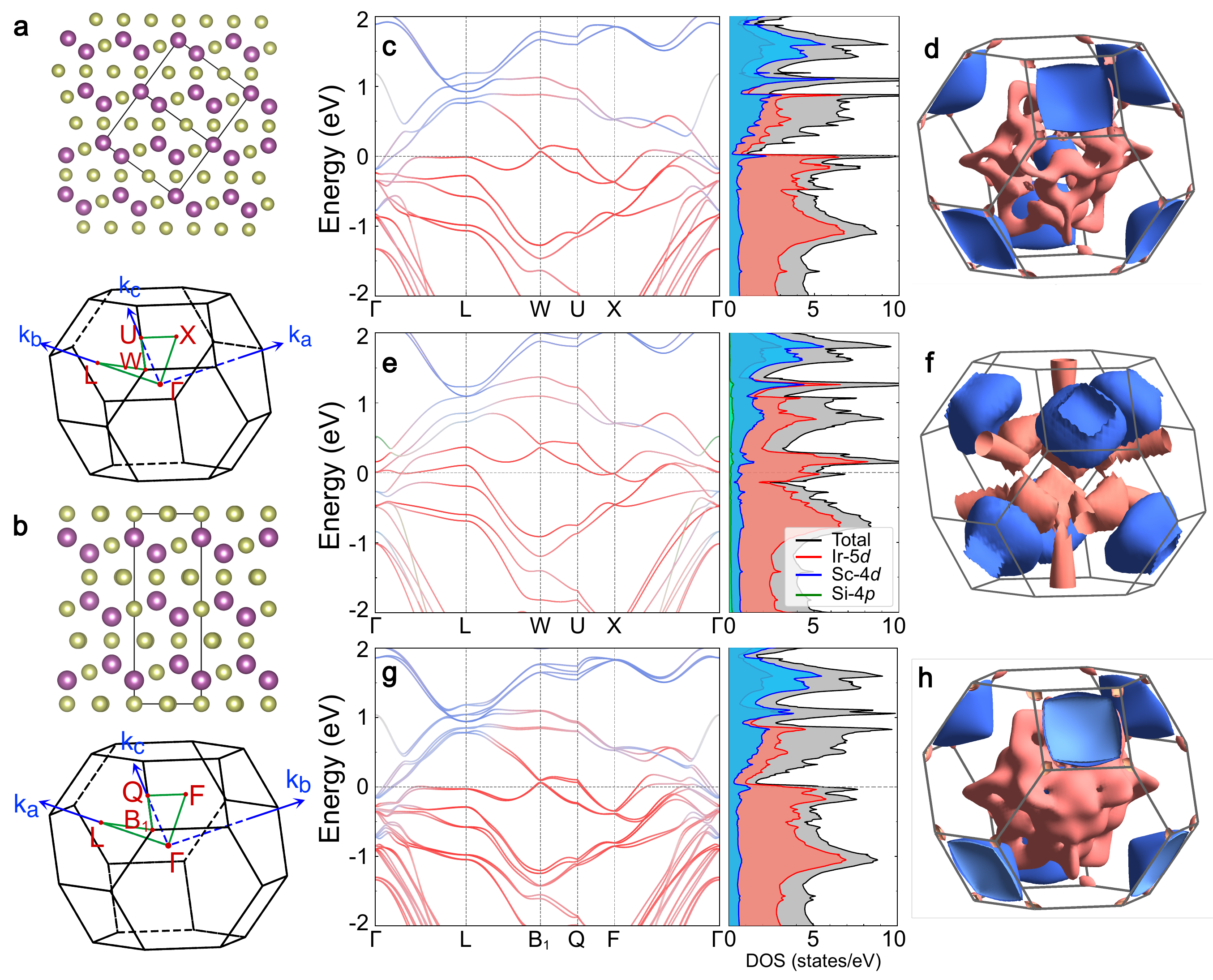}
	\caption{\label{fig_DFT} 
		Electronic band structures.
		a,b) First Brillouin zone of ScIr$_2$ with space groups $Fd\bar{3}m$ (No.\ 227) (a) and $R32$ (No.\ 155) (b).  
        c,d) The calculated electronic band structures (left panel) and DOS (right panel) (c) and Fermi surface (d) for the high-$T$ cubic phase of ScIr$_2$.
        e,f) The analogus results for the cubic ScIr$_{1.82}$Si$_{0.18}$.         
        g,h) Band structures (left panel) and DOS (right panel) (g) and Fermi surface (h) for the low-$T$ rhombohedral phase of ScIr$_2$. 
        The analogous results for ScIr$_{1.75}$Si$_{0.25}$ are shown in Figure~S3 of the Supporting Information.
		Band structures were calculated by considering the SOC. The results by ignoring the SOC are presented in Figure~S4 of the Supporting Information.}
	\end{figure*}
\clearpage

\begin{figure}[!thp]
	\centering
	\includegraphics[width=0.45\textwidth,angle= 0]{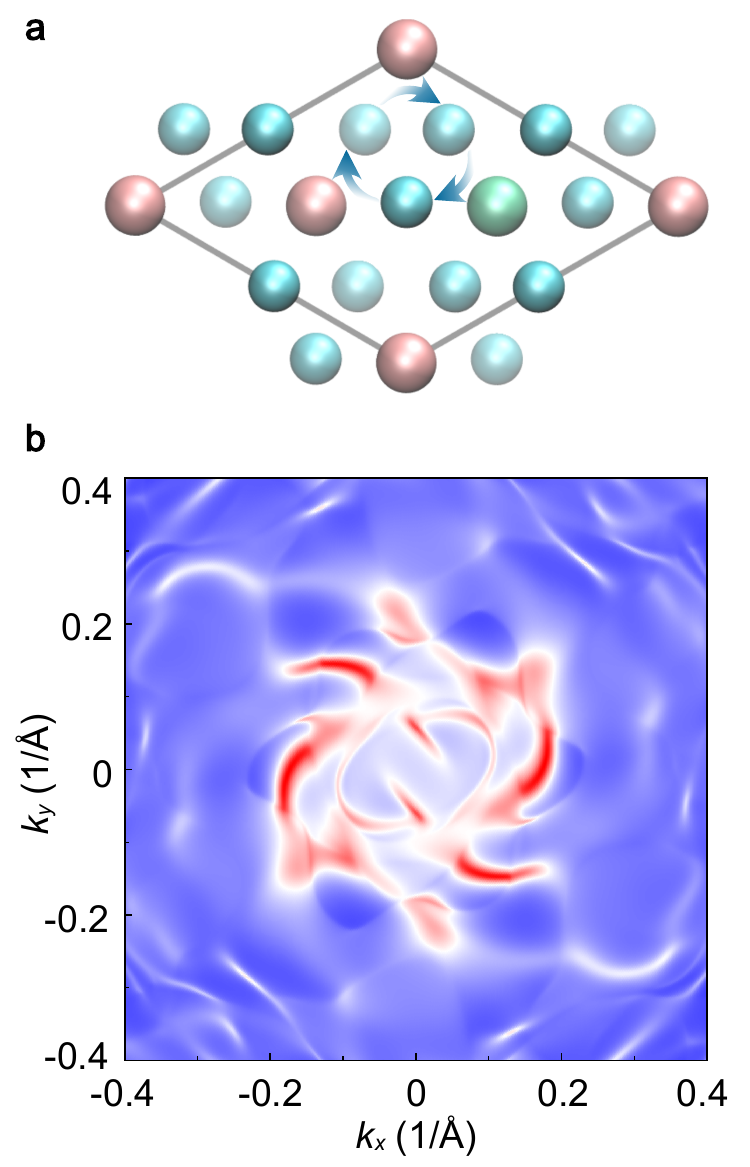}
	\caption{\label{fig:fermi_arc} Chiral crystal structure and surface states of ScIr$_2$.	
		a) Low-$T$ rhombohedral crystal structure ($R32$, No.~155) of ScIr$_2$. The Ir chiral chain (indicated by the arrows) is clearly visible in the top view along the [001]-direction.
		b) The calculated (001) surface states and surface Fermi arcs of ScIr$_2$.}
\end{figure}
\clearpage
%





\begin{table*}	
	\centering
	\caption{\label{tab:parameters} Summary of the superfluid-density and electronic specific-heat data of ScIr$_{2-x}$Si$_x$, analyzed using different models. In the gap function $g_\mathrm{k}$, $\theta$ and $\phi$ are the polar and azimuthal angles in the $k$-space. The last column lists the reduced least-squares deviations $\chi_\mathrm{r}^2$ for both the {\textmu}SR and $C_\mathrm{e}/T$ data analysis. For the two-gap model, the weight ($w$) listed in the table refers to the first $s$-wave component.}
		\vspace{3pt}
		ScIr$_2$\\
		\begin{tabular}{lcccccc}
\toprule
			\textrm{Model}&
			\textrm{$g_\mathrm{k}$ function}&
			\textrm{Gap type}&
			\textrm{$\lambda_0$ (nm)} &
			\textrm{$\Delta_0^{\text{\textmu}SR}$ (meV), $w$}&
			\textrm{$\Delta_0^{C_\mathrm{e}/T}$ (meV), $w$}&
			\textrm{$\chi^2_r$ ({\textmu}SR: $C_\mathrm{e}/T$)}\\
\toprule			
			$s$   & 1               & Nodeless   & 430(2)    & 0.26(1)     & 0.36(2) 	 & 7.0: >20 \rule{0pt}{2.6ex} \\
			$p$   & $\sin\theta$    & Point-node & 420(2)    & 0.35(1)     & 0.44(2)     & 8.2: 6.9 \\
			$d$   & $\cos2\phi$     & Line-node  & 388(3)    & 0.35(2)     & 0.39(4)     & >20: >20 \\
			$s+s$ & 1, 1            & Nodeless   & 430(2)    & 0.23(1)/0.35(1), 0.7     & 0.30(2)/0.40(2), 0.5   & 4.4: >20  \\
			$s+p$ & 1, $\sin\theta$ & Point-node & 430(2)    & 0.25(1)/0.33(1), 0.8     & 0.34(2)/0.44(2), 0.5   & 8.0: >20  \\
			$s+d$ & 1, $\cos2\phi$  & Line-node  & 425(3)    & 0.27(1)/0.35(1), 0.8     & 0.39(2)/0.40(2), 0.63  & 5.4: 3.1  \\
\bottomrule
		\end{tabular}
		\\ \vspace{10pt}		
		ScIr$_{1.82}$Si$_{0.18}$\\
		\begin{tabular}{lcccccc}
\toprule
			\textrm{Model}&
			\textrm{$g_\mathrm{k}$ function}&
			\textrm{Gap type}&
			\textrm{$\lambda_0$ (nm)} &
			\textrm{$\Delta_0^{\text{\textmu}SR}$ (meV), $w$}&
			\textrm{$\Delta_0^{C_\mathrm{e}/T}$ (meV), $w$}&
			\textrm{$\chi^2_r$ ({\textmu}SR: $C_\mathrm{e}/T$)}\\
\toprule			
			$s$   & 1                & Nodeless   & 690(3)    & 0.60(3)     & 0.46(3)    & 2.5: >20 \rule{0pt}{2.6ex} \\
			$p$   & $\sin\theta$     & Point-node & 688(3)    & 0.76(3)     & 0.58(3)    & 6.1: 16.0 \\
			$d$   & $\cos2\phi$      & Line-node  & 600(5)    & 0.73(3)     & 0.54(5)    & >20: >20  \\
			$s+s$ & 1, 1             & Nodeless   & 690(3)    & 0.49(3)/0.73(3), 0.6     & 0.44(3)/0.52(3), 0.7   & 3.1: >20  \\
			$s+p$ & 1, $\sin\theta$  & Point-node & 688(3)    & 0.60(3)/0.76(3), 0.8     & 0.41(3)/0.58(3), 0.5   & 2.4: >20  \\
			$s+d$ & 1, $\cos2\phi$   & Line-node  & 680(3)    & 0.60(3)/0.73(3), 0.8     & 0.47(3)/0.54(3), 0.6   & 2.4: 2.64 \\
\bottomrule
		\end{tabular}
\end{table*}
\clearpage

\end{document}